\documentclass[aps,pre,twocolumn,groupedaddress]{revtex4-2}

\usepackage{amsfonts,a msmath}

\usepackage{graphicx}
\usepackage{soul, color} 
\usepackage{float}

\begin{document}

\title{Robustness of community structure under edge addition}
\author{Moyi Tian}
\email[]{moyi\_tian@brown.edu}
\homepage[]{https://moyi-tian.github.io/moyi-tian}
\affiliation{Division of Applied Mathematics, Brown University, Providence, Rhode Island 02912, USA}

\author{Pablo Moriano}
\email[]{moriano@ornl.gov}
\homepage[]{https://pmoriano.com}
\affiliation{Computer Science and Mathematics Division, Oak Ridge National Laboratory, Oak Ridge, Tennessee 37830, USA}

\date{\today}

\begin{abstract}
Communities often represent key structural and functional clusters in networks. To preserve such communities, it is important to understand their robustness under network perturbations. Previous work in community robustness analysis has focused on studying changes in the community structure as a response of edge rewiring and node or edge removal. However, the impact of increasing connectivity on the robustness of communities in networked systems is relatively unexplored. Studying the limits of community robustness under edge addition is crucial to better understanding the cases in which density expands or false edges erroneously appear. In this paper, we analyze the effect of edge addition on community robustness in synthetic and empirical temporal networks. We study two scenarios of edge addition: random and targeted. We use four community detection algorithms, Infomap, Label Propagation, Leiden, and Louvain, and demonstrate the results in community similarity metrics. The experiments on synthetic networks show that communities are more robust when the initial partition is stronger or the edge addition is random, and the experiments on empirical data also indicate that robustness performance can be affected by the community similarity metric. Overall, our results suggest that the communities identified by the different types of community detection algorithms exhibit different levels of robustness, and so the robustness of communities depends strongly on the choice of detection method.
\end{abstract}

\maketitle


\section{\label{sec:level1} Introduction}

Many complex systems, such as critical infrastructures, biological networks and social groups, exhibit network structures consisting of nodes and edges that capture their connectivity \cite{ComplexNetworks}. Extracting different features from these networks is useful to better understand their structure and function \cite{NetworkStructure, GNBench}. Among different properties of networks, many real networked systems demonstrate community structures, or clusters, which are groups of nodes that have higher probability of sharing links with other nodes within the same group than with nodes in different groups. These communities typically represent essential functional or behavioral units in the networks \cite{GNBench, Detection, Detection_UserGuide}. Therefore, it is important to understand the conditions under which they persist. Community \textit{robustness} describes how much network perturbation a community structure can tolerate while still being able to recover its original structure \mbox{\cite{Detection, robustKLN}}. The community robustness problem is of practical interest because we want to understand how well the components in networks can sustain their basic functionalities when facing errors or attacks \cite[Chapter~8]{barabasi_network}. Usually, network perturbations are used to model either random failures or deliberate attacks in real networked systems \cite{RobustnessReview}.

Studying the robustness of communities in networks confronts two main challenges. First, although community structure in networks can provide insights into the organization of nodes based on their connectivity, there is no universal definition of community. This makes the detection of communities itself an ill-posed problem, and hence, in practice, discovering community structure often depends on the methods and application \cite{Detection_UserGuide}. Second, community robustness involves network perturbations, and there are various ways a network can be modified. Studies so far have focused on edge rewiring and node or edge removal as the network perturbation schemes for studying a system's ability to sustain basic functionality when some components fail~\cite{robustKLN, edge_attack, node_removal, robust_node_rm, node_removal_edge_rewiring, robust_validation, improve_robust}.

However, real networked systems, such as communication, citation, or social networks, often increase connectivity over time, so it is important to consider the scenario of an increasing number of edges \cite{DensifyingGraphs}. Added edges can also simulate errors or attacks \cite{node_removal, attack, addition_error}. For example, when simulating errors, added edges can simulate false-positive edges (i.e., edges that are present in data erroneously but do not actually represent the real relations). This is one type of measurement error commonly found, such as in online community, communication, and collaboration network data \cite{measurement_error}. Similarly, added edges can also simulate deliberate attacks intended to destroy the established community structure to quickly disrupt the functionality of the system as in the case of a Distributed Denial of Service attack \cite{DoS1, DoS2}. Therefore, there is a need to understand the effect of network densification on community robustness. 

Here, we conduct a systematic study to understand the impact of edge addition on the robustness of communities. We focus on both the synthetic Lancichinetti-Fortunato-Radicchi (LFR) benchmark graphs \cite{LFRBenchNew} and empirical networks to investigate the limits of the robustness of the initial community structure as the network is perturbed through edge addition. We study two scenarios of edge addition, namely random and targeted addition. Random addition selects from the set of all nonexistent edges, and this process is analogous to a random error in the system \cite{measurement_error, addition_error}. Targeted addition selects only from the nonexistent, cross-community edges, which we propose to use to simulate attacks. We compute the effects on community robustness by using Normalized Mutual Information (NMI) \cite{NMI}, which is a community similarity measure often used in network community analysis, and by using element-centric clustering similarity \cite{element-centric} to control for biases when comparing clusters. We specifically select four community detection algorithms commonly used in community detection benchmarking studies \cite{ClusterAlg, AnaCluQua}: Infomap, Label Propagation, Leiden, and Louvain. We also demonstrate and compare community robustness results by using the same set of community detection algorithms on empirical email network data, in which the edge density increases over time.

Our results suggest that the chosen clustering algorithm strongly affects community robustness under edge addition. Specifically, in both synthetic and empirical networks, we observe that for different types of community detection algorithms, the similarity measures between communities in the original and the perturbed networks decay with distinct rates while more and more edges are being added. Additionally, in synthetic experiments on LFR benchmark graphs, we observe that with a smaller mixing parameter, which means that initial communities are more loosely connected to each other, the communities tend to be more robust. Synthetic experimental results also align with the expectation that targeted edge addition tends to destroy the original community structure more rapidly compared with random addition, so communities are less robust with targeted addition. In experiments that use empirical data, we observe that the choice of community similarity metrics, NMI or element-centric clustering similarity in particular, also affects the results on community robustness.

\section{\label{sec:level3} Methods}

\subsection{LFR benchmark} 

The LFR benchmark graphs \cite{LFRBench, LFRBenchNew} are commonly used in network community studies to create graphs with ground-truth partitions. The advantage of using an LFR benchmark is that the degree and the community size both follow power-law distributions, which more closely resemble the properties observed in many real-world networks \cite{LFRBench, NetworkStructure, power_law, power_law2, power_law3}. The exponents of degree distributions are controlled by parameter $\alpha$, and the exponents of community size distributions are controlled by parameter $\beta$. We take the typical values of the exponents observed in real networks, i.e., $\alpha=-2$ and $\beta=-1$ \cite{LFRBench}. Other parameters required for generating LFR benchmark graphs include the average degree $\langle k \rangle$, maximum degree $maxk$, minimum community size $minc$, maximum community size $maxc$, and the mixing parameter $\mu$, which represents the fraction of nodes that each node shares edges with across different communities. The lower the $\mu$, the higher the ratio between the number of internal and external connections. This leads to higher modularity, which is a quality function commonly used to express the strength of communities in studies of communities \cite{Detection}. Thus, the smaller the $\mu$, the stronger the partition.

Although we use LFR benchmark graphs for the synthetic experiments, our method is generally applicable to any type of benchmark graph that provides ground-truth community labels, such as the Girvan-Newman benchmark \cite{GNBench} and the stochastic block model \cite{SBM_original}. We generate the LFR benchmark graphs by using the publicly available implementation \cite{Fortunato_website} of the algorithm described in Ref. \cite{LFRBenchNew}. We conduct experiments on $1000$ nodes and then on $10\,000$ nodes to study the effect of perturbations at different scales. We also examine the effect of the strength of the initial community by varying the mixing parameter, $\mu$. The specific parameter values used for generating the LFR benchmark graphs are listed in Table~\ref{tab:table1}.

\begin{table*}[ht]
\caption{\label{tab:table1}%
Synthetic experiment parameters.
}
\begin{ruledtabular}
\begin{tabular}{clc}
\textrm{\textbf{Parameter}}&
\textrm{\textbf{Description}}&
\textrm{\textbf{Value}}\\
\hline
$N$ & Number of nodes & $1000, 10\,000$\\
$maxk$ & Max degree for LFR& $0.1 N$\\
$\langle k \rangle$ & Average degree for LFR& $25$\\
$minc$ & Min community size  for LFR& $50$\\
$maxc$ & Max community size  for LFR& $0.1 N$\\
$\alpha$ & Degree distribution exponent  for LFR& $-2$\\
$\beta$ & Community size distribution exponent for LFR& $-1$\\
$\mu$ & Mixing parameter for LFR& $0.01, 0.1, 0.2, 0.3, 0.4, 0.5$\\
$h$ & Times of initial number of edges added up to & $10$\\
$s$ & Number of steps to add edges& $50$\\
$r$ & Number of realizations per step & $50$\\
\end{tabular}
\end{ruledtabular}
\end{table*}

\subsection{Community detection}

Community detection is the task of assigning nodes in a network into clusters based on topological similarity \cite{Detection}. Nodes within the same community are more densely connected compared with the ones across distinct communities. Various community detection algorithms have been developed to find clusters in networks. These algorithms are based on different methodologies to achieve their optimal clustering solutions. In this work, we use four algorithms based on three popular methods of community detection: information-theoretic-based algorithm Infomap \cite{infomap}, message-passing-based algorithm Label Propagation \cite{lpm}, and modularity-based algorithms Leiden (partly based on smart local move algorithm and improved from Louvain) \cite{leiden} and Louvain \cite{louvain}. These chosen algorithms have low enough computational complexity \cite{detection_analysis, leiden} to accomplish our experiments in reasonable run time. For Infomap, Label Propagation, and Louvain, we use the publicly available package \cite{Fortunato_website} implemented by \citet{ConsensusClu}. The package for Leiden is publicly available on GitHub \cite{leiden_website} and described in the original paper \cite{leiden}. We use the undirected and unweighted implementations for all four algorithms to ensure consistency with the LFR benchmark graphs we generate. Rigorous analysis of these four clustering algorithms is presented in previous work \cite{AnaCluQua}.

The community detection algorithm is an essential component for studying communities because, although the benchmark graphs have ground-truth labels of the communities, few empirical networks have ground-truth partitions. Moreover, the temporal evolution of the networks makes it more difficult to obtain ground-truth clustering information at all times. Notably, although graph embeddings have become popular for downstream tasks, they have not been developed thoroughly enough to efficiently achieve good graph clustering results; their hyper-parameter tuning is cumbersome when attempting good performance \cite{CommDetect_Embedding}. By comparison, traditional community detection algorithms require no parameter tuning but can provide relatively good clustering results in a reasonable run time. For these reasons, we use the four well-developed community detection algorithms for our experiments.

\subsection{\label{sec:methods_metrics} Community similarity}

To measure similarity between communities, we use NMI \cite{NMI} and element-centric clustering similarity \cite{element-centric} as the metrics. Without loss of generality, suppose that $C_1$ and $C_2$ are partitions on the same set of $N$ nodes. The NMI score of the two partitions is defined as

\begin{equation}
NMI (C_1, C_2) = \frac{ -2 \displaystyle \sum_{i = 1}^{|C_1|} \sum_{j = 1}^{|C_2|} \mathbb{P} (i, j) \log \left(\frac{\mathbb{P} (i, j)}{\mathbb{P}_1 (i) \mathbb{P}_2 (j)} \right) }{ \displaystyle \sum_{i = 1}^{|C_1|} \mathbb{P}_1 (i) \log \mathbb{P}_1 (i) + \displaystyle \sum_{j = 1}^{|C_2|} \mathbb{P}_2 (j) \log \mathbb{P}_2 (j) }{,}
\label{eq:NMI}
\end{equation}
where $\mathbb{P}_1 (i) = \frac{|{C_{1}}_{i}|}{N}$, $\mathbb{P}_2 (j) = \frac{|{C_{2}}_{j}|}{N}$, and \mbox{$\mathbb{P} (i,j) = \frac{|{C_{1}}_{i} \cap {C_{2}}_{j}|}{N}$} for ${C_{1}}_{i} \in C_1$, ${C_{2}}_{j} \in C_2$ as the clusters of partition $C_1$ and $C_2$, respectively. We use the scikit-learn implementation of NMI \cite{scikit-learn} in our experiments.

In addition to NMI, we also compute the similarity between communities by using element-centric clustering similarity, which better copes with issues such as bias in randomized membership, bias in skewed cluster sizes, and the problem of matching \cite{element-centric}. Although NMI tends to favor more clusters, element-centric clustering similarity overcomes such bias in the number of clusters. We report the experimental results computed in this metric using the CluSim package \cite{clusim} along with the default parameter. Both NMI and element-centric clustering similarity range from $0$ to $1$, where a higher value means more similarity between partitions.

\subsection{Experimental procedure}
We experiment on several computer-generated networks and empirical temporal networks to examine the robustness of their community structures. The empirical networks are real-world data, and we have no control over their intrinsic network properties. The ground-truth communities and the interpretability of the clusterings found by detection algorithms on these data are often unclear \cite{GroundTruth}, so in the studies on clusterings in networks, synthetic networks often serve as handy examples for tests. Here, we illustrate the details of the experimental procedures for synthetic and empirical networks. The tests on synthetic and empirical networks are designed differently because the synthetic network is stationary without natural perturbations, whereas the empirical data does not have ground-truth community labels and requires further cleaning to work as comparable examples. Our code is publicly available at: \href{http://github.com/Moyi-Tian/CommunityRobustness}{http://github.com/Moyi-Tian/CommunityRobustness}.

\subsubsection{\label{subsec:methods_exp_synthetic} Experiments on synthetic networks}

Suppose the initial network is $G = (V,E)$, where $V = \{1,2, \ldots,N \}$ is the set of $N$ nodes, and $E = \{ e_{ij}: i,j \in V\}$ is the set of $M$ edges. Let $E^c$ be the set of edges in the complement graph of $G$. The benchmark graph provides each node a community label denoted by $c_i$ for $i \in V$. The graph partition is then $C = \bigcup_{k \in \bigcup_{i \in V} \{ c_i \}} \left\{ \bigcup_{j \in V} \{j: c_j = k \} \right\}$. We also define a set for nonexistent edges across different communities denoted by $E_{inter}^c = \{ e_{ij} \in E^c: c_i \ne c_j, i,j \in V \}$. To start, we choose a community detection algorithm and specify parameters $h$, $s$, and $r \in \mathbb{Z}_+$, where $h$ is how many times the initial number of edges is added up, $s$ is the number of steps to add edges, and $r$ is the number of realizations per step. The parameters we use for the synthetic experiments are listed in Table~\ref{tab:table1}. 

We illustrate the effects of edge addition on community structures through two different network perturbations: random addition and targeted addition. Let $t \in \{0,1,2,\ldots,s \}$. For the random addition approach, we select $E_{\nu} \subseteq E^c$ uniformly at random. For targeted addition, we choose $E_{\nu} \subseteq E_{inter}^c$ uniformly at random, where $|E_{\nu}| = \left\lfloor \frac{hM}{s} \right\rfloor t$. Notably, the random addition is analogous to the case in which random errors or random additional connections appear. Conversely, targeted addition simulates the case in which much of the information about the community structure is known, and there is an intention to break the current partitions through increasing connectivity. In each of these edge-addition configurations, we create a new graph, $G' = (V, E')$, such that $E' = E \cup E_{\nu}$.

We then apply the specific community detection algorithm on $G'$ to yield an associated graph partition, $C'$. We repeat the previous steps for $r$-independent times at each $t$: specifically $r$ realizations at every single step $t$ over the total $s$ steps. Suppose the chosen community similarity metric is $\mathcal{S}$. We then calculate the metric score $\mathcal{S}_k (C, C'_k)$ for each realization, $k$, where $C'_k$ is the associated graph partition, and report the average $\mathcal{S}_{avg} = \frac{1}{r} \sum_{k = 1}^{r} \mathcal{S}_k (C, C'_k)$.

\subsubsection{\label{subsec:methods_exp_empirical} Experiments on empirical temporal networks}

Temporal email networks are natural candidates for empirical experiments and are comparable to the synthetic experiments. The email conversations between users in these networks naturally emerge at their sent time, which can be directly considered as additional edges over time steps.

Unlike the benchmark graphs, empirical networks do not have ground-truth community labels, so we must first identify a reliable community partition on the initial graph before the start of network perturbation. In doing so, we apply the fast consensus algorithm \cite{fast_consensus}, which is based on the idea of consensus clustering \cite{ConsensusClu}. Because established community detection algorithms produce nondeterministic partitions, consensus clustering was proposed for more stable and accurate results after iterations among multiple clustering results given a prespecified clustering method. There are two parameters we must specify as inputs for the fast consensus algorithm: the number of partitions, $np$, and the clustering method. The default value for $np$ in the fast consensus algorithm is set to be $20$, which is also the value used for tests demonstrated in the original paper \cite{fast_consensus}. We also found that $np = 20$ balances performance and run time. This choice of the value is specified in Table~\ref{tab:table2}. For the clustering method, we choose it to be consistent with the one used for community detection in the later perturbed networks.

In addition, there are other issues that must be addressed before testing on email networks with time stamps associated with edge emergence. 

First, the problem is finding appropriate empirical network examples on which we can perform experiments comparable to the synthetic case. Specifically, the empirical network should have a temporally growing number of edges on a fixed set of nodes, and the growth within the recorded time frame should be on the comparable scale as the synthetic experiment, where we add edges up to $10\times$. The problem is that temporal email networks always expand in the number of edges \textit{and} in the number of nodes. If we look at the graph on the set of all users who appear within the given time frame, then we usually find that only a tiny fraction of edges, or sometimes none, are added until the end of the recorded time. This is because there are always new users joining the networks, occasionally even at the very end. For some networks, there is also a co-occurring issue that many users are not very active and do not contribute many new communications (i.e., edges) over time. This is why, rather than using the entire network datasets as obtained, we instead first identify appropriate subnetworks extracted from the original email data and then use them as examples for our empirical experiments. There are different ways to select subnetworks from the original ones. When the entire network is small enough, choosing subnetworks based on an exhaustive search may be possible. However, due to the size of our original empirical datasets, checking all possible combinations of nodes and the growth in density of their induced subnetworks will be computationally expensive. Therefore, our approach is to search over a family of subnetworks induced by the first $n$ nodes showing up in time with $n$ swept from $1$ to $N$ where $N$ is the total number of users present in the full dataset. We then look at the growth in density for each of these subnetworks and select an appropriate one as the example to use. More details of our subnetwork selection and the corresponding network properties are described in Sec.~\ref{sec:empirical_results}.

Second, email networks usually have multiedges emerging in time because several emails can be sent between the same pair of users, but the synthetic networks we test on have no multiedges. Here, we align our empirical experiment with the synthetic case by preprocessing the network data so that there are no repeating edges. We do so by removing the edges that arrived later and already showed up once at a previous time from the edge list. In this way, we consider an edge to represent an existing relationship between users, thereby omitting the number of conversations that occurred. 

Third, we must also identify which network should be treated as the initial network. The empirical networks always start with zero edges, but it is not meaningful to use the null graph because then no communities will ever exist at the initial time. Also, the community detection algorithms take in only the edge lists and so only the nodes incident to the edges present in the list are assigned with community labels. So, to use the algorithms, we must ensure all nodes appear in the edge lists. Therefore, in our empirical experiment, we choose the initial network by picking the first one without any isolated nodes when growing the network in time. This is achieved by adding edges one at a time according to their associated time stamp and checking at which time every node is at least incident to one edge.

In our empirical experiments, we use $np$ for fast consensus, and we also have parameters $s$ and $r$. Here $s$ refers to the number of steps with respect to time, and $r$ is the number of realizations per step. Table~\ref{tab:table2} lists the parameter values for the empirical experiments.

\vspace*{-10pt}
\begin{table}[!ht]
\caption{\label{tab:table2}%
Empirical experiment parameters.
}
\begin{ruledtabular}
\begin{tabular}{clc}
\textrm{\textbf{Parameter}}&
\textrm{\textbf{Description}}&
\textrm{\textbf{Value}}\\
\hline
$np$ & Number of partitions (fast consensus) & $20$\\
$s$ & Number of steps& $50$\\
$r$ & Number of realizations per step & $10$\\
\end{tabular}
\end{ruledtabular}
\end{table}
\vspace*{-2pt}

As previously mentioned, we preprocess and identify suitable empirical networks for our experiment. The experimental procedure is as follows. Suppose that we have a precleaned empirical network dataset (without multiedge) with $V = \{1,2,\ldots, N \}$ to be the set of nodes and $M$ to be the number of edges. Recall that edges appear in time one by one, so there are $M$ associated time stamps. At time stamp $i \in \{1,2,\ldots, M \}$, we denote the emergent edge pointing from node ${v_s}(i)$ to node ${v_t}(i)$ by $e_{{v_s}(i),{v_t}(i)}$, where ${v_s}(i), {v_t}(i) \in V$. Using these notations, the dataset can be presented as an edge list $\{e_{{v_s}(1),{v_t}(1)}, e_{{v_s}(2),{v_t}(2)}, \ldots, e_{{v_s}(i),{v_t}(i)}, \ldots, e_{{v_s}(M),{v_t}(M)}\}$. Then we grow the network until the time stamp \mbox{$t_{0} = \min \left\{ t_s: \bigcup_{1 \le i \le t_s} \left\{ {v_s}(i),{v_t}(i) \right\} = V, 1 \le t_s \le M \right\}$}, which is the first time when there are no isolated nodes. The initial network is chosen to be $G_{0} = (V, E_{0})$, where $E_{0} = \{e_{{v_s}(i),{v_t}(i)}: 1 \le i \le t_{0} \}$. Before each experiment, we specify a clustering method and parameters, $np, s, r \in \mathbb{Z}_+$ ($s$ is upper bounded by $M - t_{0}$, and our preprocessing should provide an appropriate dataset that $M \gg t_{0}$). We first apply fast consensus on $G_{0}$ using $np$ as the hyperparameter for the number of partitions and the chosen method for the clustering algorithm. The consensus algorithm yields $np$ initial partitions $C_{0, w}$ for $1 \le w \le np$. Then we simulate on the evolved networks over time. Specifically, for $1 \le p \le s$, let $t_p = t_{0} + \left\lfloor \frac{M - t_{0}}{s} \right\rfloor p$. The new graph $G_p = (V, E_p)$ at time step $p$ is constructed such that $E_p = \{e_{{v_{s}}(i),{v_{t}}(i)}: 1 \le i \le t_{p} \}$. Use the chosen clustering method to find partitions of $G_p$ for $r$ times independently and denote the corresponding clustering results as $C_{p,q}$ for $1 \le q \le r$. For each $1 \le p \le s$, we report the average community similarity metric score: $\mathcal{S}_{avg,p} = \frac{1}{np \cdot r} \sum_{w = 1}^{np} \sum_{q = 1}^{r} \mathcal{S}(C_{0, w}, C_{p,q})$.

\section{\label{sec:level4} Results and Discussion}

This section describes the computational results from the synthetic experiments on LFR benchmark graphs and the empirical experiments on subnetworks obtained from temporal email networks. 

Note that while more edges are added to the initial network, it is likely that the community structure evolves over perturbation. However, our focus is not on tracking changes in the community structure itself but on understanding the limits of the robustness of the initial community structure under edge-addition perturbation.

\subsection{Synthetic networks}

We test on LFR benchmark graphs with $1000$ and $10\,000$ nodes and report the results calculated with the NMI metric. The results for the same series of experiments using the element-centric clustering similarity metric are presented in Appendix~\ref{app:syn-clusim}. We also include the associated plots of standard deviation for all synthetic experiments in Appendix~\ref{app:syn-std}.

\begin{figure}[!htbp]
\includegraphics[width=8.6cm]{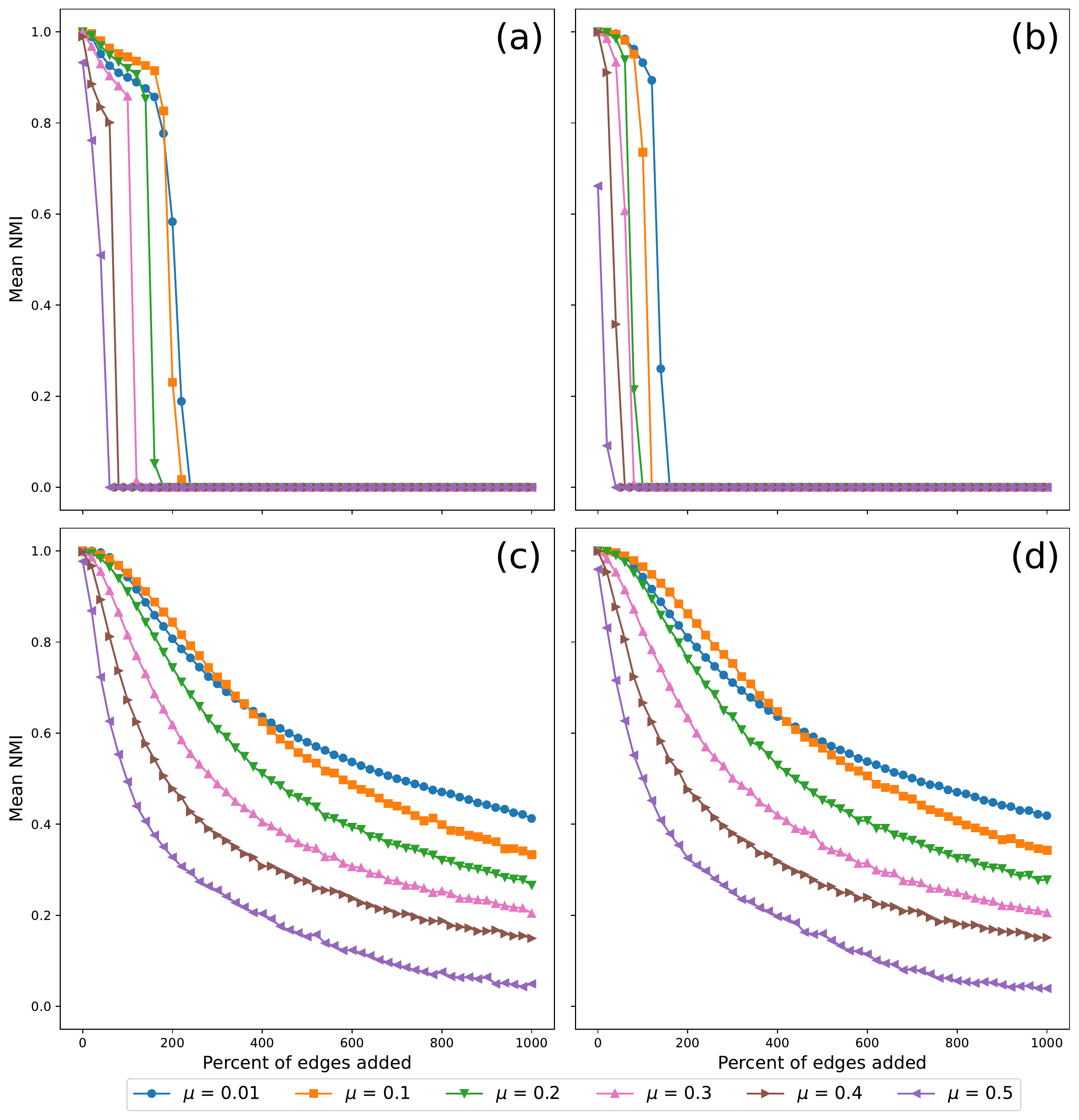}
\caption{\label{fig:LFR10000} Mean NMI over the percentage of edges added uniformly at random on LFR benchmark graphs with $10\,000$ nodes. Communities detected by (a) Infomap, (b) Label Propagation, (c) Leiden, and (d) Louvain. Same parameter values are used as with the $1000$-node case. The maximum amount of edges we add is about $2.5\%$ of all nonexistent edges.}
\end{figure}

Figures~\ref{fig:LFR1000} and~\ref{fig:LFR10000} show how the four different community detection algorithms---Infomap, Label Propagation, Leiden, and Louvain---perform when the LFR benchmarks with $1000$ and $10\,000$ nodes are perturbed under uniformly random edge addition. Recall that in this setting, the edges added at each step are selected uniformly at random from all the nonexistent edges in the initial network, while multiedges are prohibited. This is analogous to random errors in real-world networks.

\begin{figure}[!htbp]
\includegraphics[width=8.6cm]{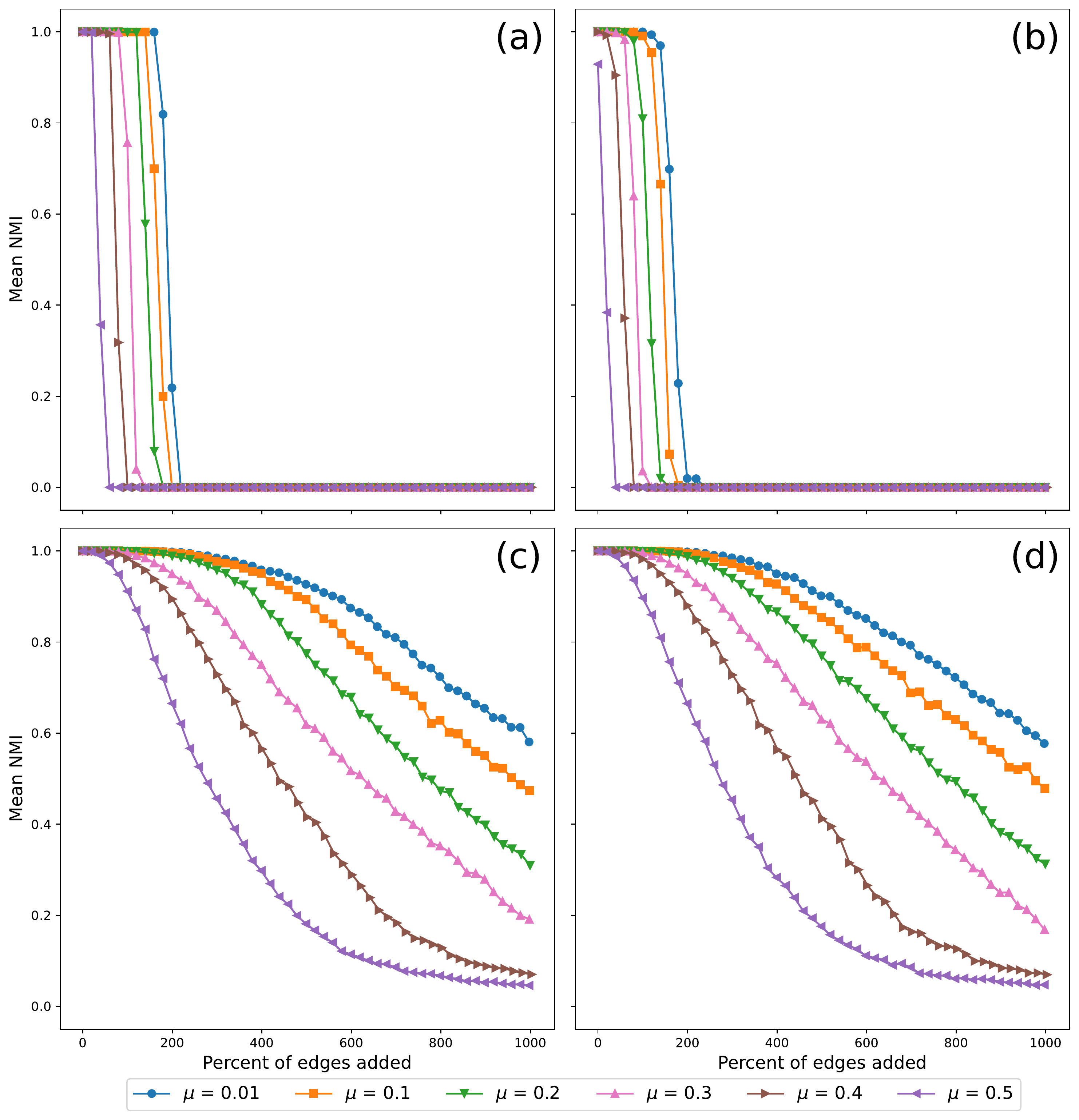}
\caption{\label{fig:LFR1000} Mean NMI over the percentage of edges added uniformly at random on LFR benchmark graphs with $1000$ nodes. Communities detected by (a) Infomap, (b) Label Propagation, (c) Leiden, and (d) Louvain. Average degree is 25. Edges are added up to $10\times$ the original number over 50 independent steps and are selected without producing multiedges. For each algorithm, we average NMI between the ground-truth partition and new partitions of the perturbed network over 50 independent runs at each step. The maximum amount of edges we add is about $25\%$ of all nonexistent edges.}
\end{figure}

LFR benchmark graphs with lower $\mu$ values (i.e., stronger initial partitions) have higher NMI values compared with the ones with higher $\mu$ after $10\times$ the original number of edges are added. Note that adding edges is essentially a process of balancing the fraction of intercommunity edges with that of intracommunity edges. Hence, this observation aligns with our intuition because when a community partition is stronger, it requires more edges to be added until the established community structure becomes less clear, meaning that the community structure is more robust. 

The modularity-based algorithms, Louvain and Leiden, have relatively higher NMI scores vs Infomap and Label Propagation. For example, if we look at the $\mu = 0.3$ curves in Fig.~\ref{fig:LFR1000}, then Louvain and Leiden need about $6.3\times$ the original number of edges to be added until the similarity scores drop $50\%$, whereas Infomap and Label Propagation only need about $1.1\times$ and $0.9\times$, respectively. This indicates that Louvain and Leiden are better at detecting community structures that are similar to those initial ones in the LFR benchmark graphs after a large number of edges is added. The exact reasons are unclear, but a plausible explanation is that the algorithms have different behaviors due to their assumptions and methodologies when performed on graphs with different intrinsic network properties. Specifically, we observe that our method of appending edges shifts the degree distribution from the initial power-law distribution to a distribution closer to the binomial. 

In our experiments, Infomap and Label Propagation end up only finding one giant community for the entire network after about $1\times$ or $2\times$ the original number of edges are added. According to the original Infomap paper \cite{infomap}, this flow-based method excels at identifying movement patterns, whereas the modularity-based method is better at detecting structure in networks with pairwise relations but not many flows. The rapid drop in NMI for Infomap could result from our perturbation methods focusing on adding connections between nodes because this focus more directly alters the topological structure of the graph rather than representing any flow of patterns. For Label Propagation \cite{lpm}, the authors explicitly state that their method only detects a single community for the giant connected component in those homogeneous networks without community structures, such as the Erd\H{o}s-R\'{e}nyi model. A reason for the rapid drop in NMI for Label Propagation could be that as we fix the number of nodes and keep adding edges selected uniformly at random among the nonexistent edges (with restriction to the cross-community ones for the targeted case), the perturbed graph gradually becomes more and more homogeneous, which gets closer to the structure of an Erd\H{o}s-R\'{e}nyi random graph while growing in its density.

\begin{figure}[!htbp]
\includegraphics[width=8.6cm]{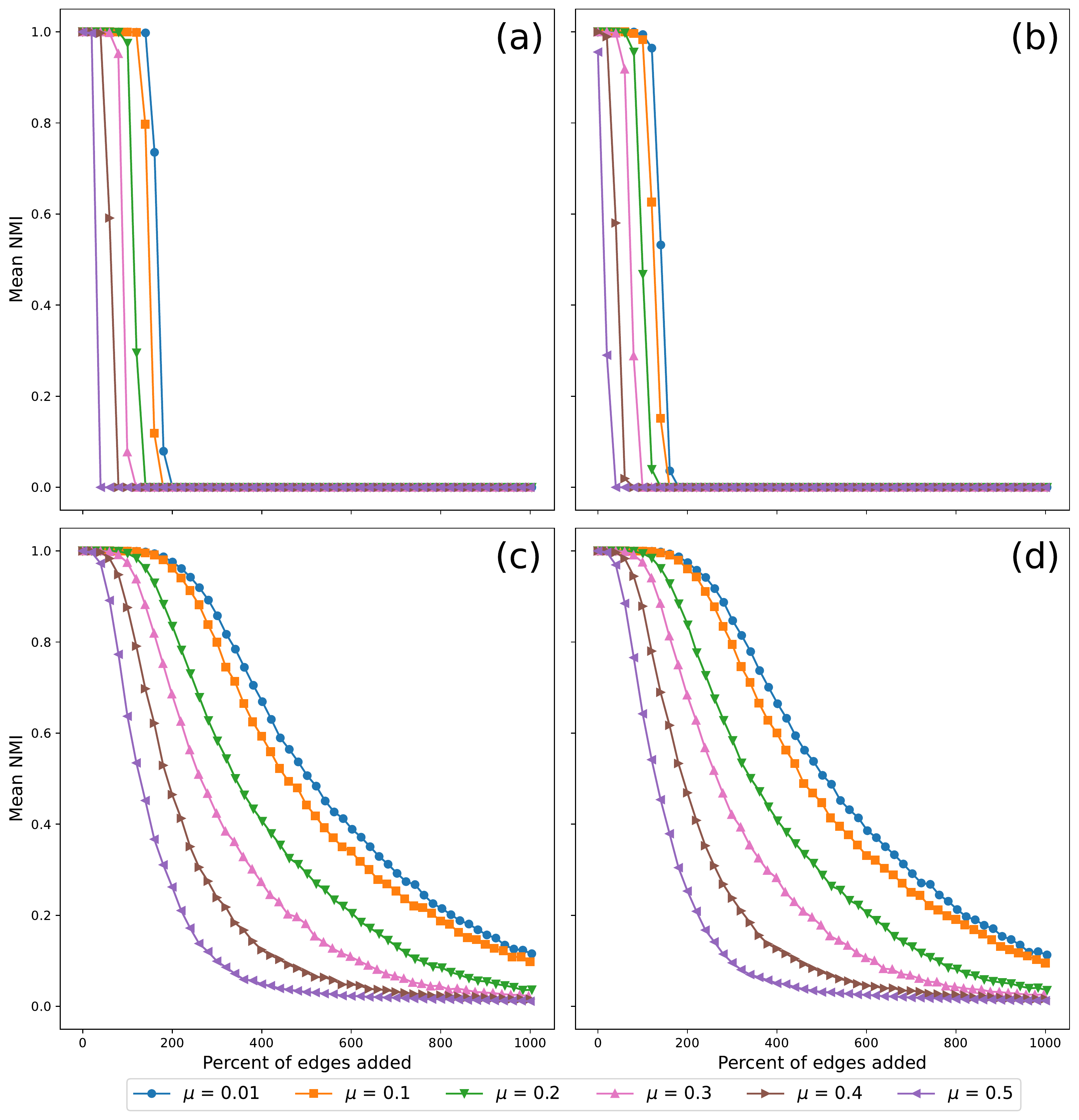}
\caption{\label{fig:LFR1000_target} Mean NMI over the percentage of edges added that are selected uniformly at random across different communities on LFR benchmark graphs with $1000$ nodes. Communities detected by (a) Infomap, (b) Label Propagation, (c) Leiden, and (d) Louvain. Same parameter values are used as in previous experiments. The maximum numbers of edges we add are between $26\%$ and $28\%$ of all nonexistent cross-community edges for all $\mu$.}
\end{figure}

\begin{figure}[!htbp]
\includegraphics[width=8.6cm]{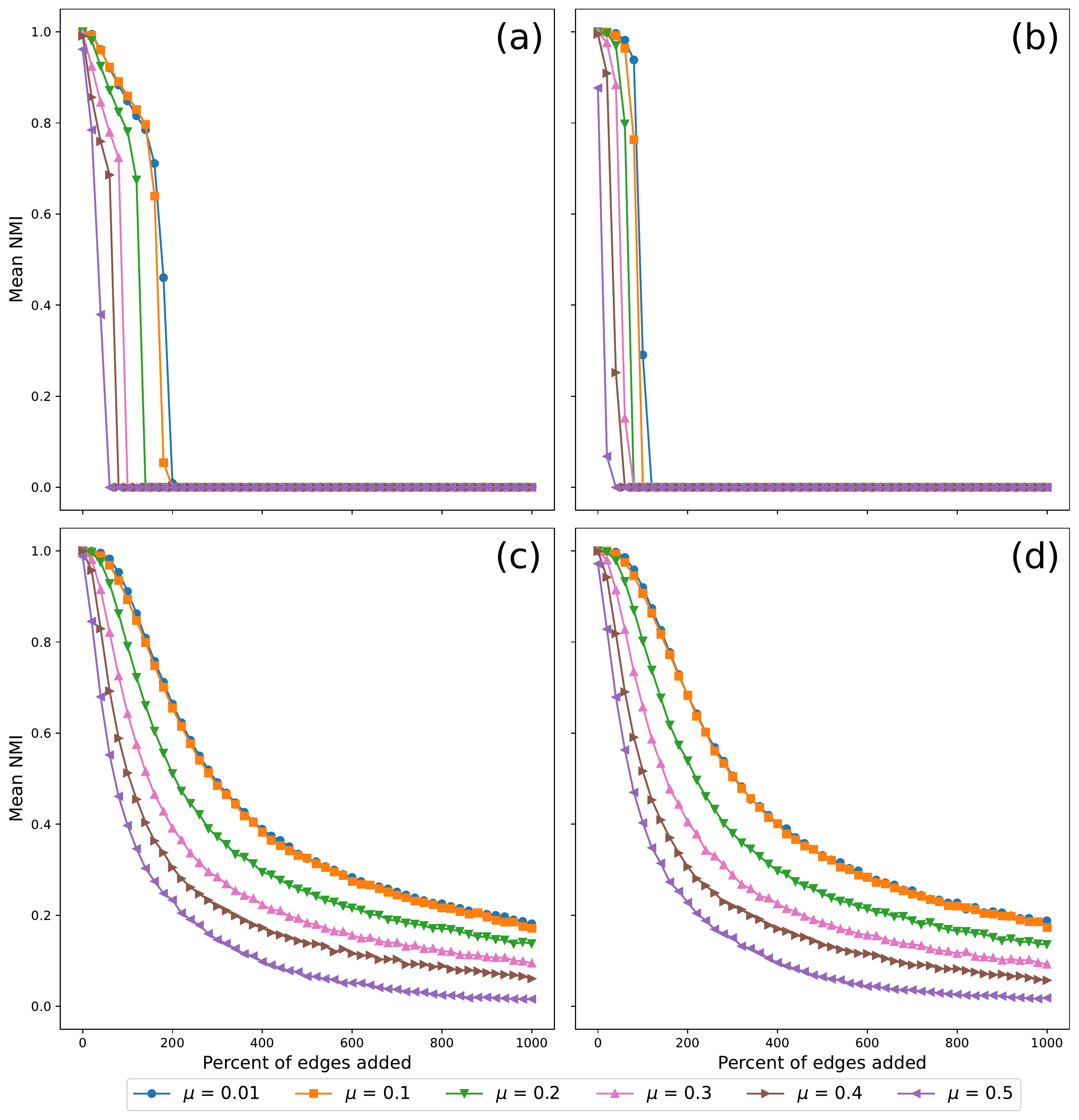}
\caption{\label{fig:LFR10000_target} Mean NMI over the percentage of edges added that are selected uniformly at random across different communities on LFR benchmark graphs with $10\,000$ nodes. Communities detected by (a) Infomap, (b) Label Propagation, (c) Leiden, and (d) Louvain. Same parameter values are used as in previous experiments. The maximum numbers of edges we add are in the range from $2.4\%$ and $2.7\%$ of all nonexistent cross-community edges for all $\mu$.}
\end{figure}

Figures~\ref{fig:LFR1000_target} and~\ref{fig:LFR10000_target} demonstrate the results on LFR benchmark graphs under targeted edge addition for $1000$ nodes and $10\,000$ nodes, respectively, which means that we restrict the new edges to be across distinct communities in the initial networks. Because the appended edges are forced to connect different communities, the targeted addition should be able to destroy the original community structure quicker than random addition, which is similar to the purpose of attacks on real networks. As expected, the targeted edge addition has relatively lower NMI values for all four algorithms vs the previous results with random edge addition. For example, if we again look at the $\mu = 0.3$ curves with $1000$ nodes in Fig.~\ref{fig:LFR1000_target} but for targeted addition, Louvain and Leiden need to add about $2.7\times$ (vs $6.3\times$ for random addition), whereas Infomap and Label Propagation need to add about $0.9\times$ and $0.7\times$ (vs $1.1\times$ and $0.9\times$ for random addition) the original number of edges, respectively, to drop the similarity scores below $0.5$. In targeted addition, we also observe that networks with stronger initial community structures tend to be more robust, and this is the same trend we see in random addition. Moreover, we again observe that Louvain and Leiden are better at detecting clusters similar to the originals.

Notably, in experiments with larger networks, specifically in Fig.~\ref{fig:LFR10000}, the curves for $\mu = 0.01$ and $\mu = 0.1$ with Louvain and Leiden cross each other when about $4\times$ of the original number of edges is added. Also, in Fig.~\ref{fig:LFR10000_target}, the curves with Louvain and Leiden for $\mu = 0.01$ and $\mu = 0.1$ are almost superimposed on each other. Recall that, as we previously discuss in Sec.~\ref{sec:methods_metrics}, there are limitations with the NMI metric. For this reason, we also compute results with the element-centric clustering similarity, for which there is a clear separation between curves. We show these results in Figs.~\ref{fig:LFR10000_elsim} and~\ref{fig:LFR10000_elsim_target} in Appendix \ref{app:syn-clusim}.

\subsection{\label{sec:empirical_results} Empirical networks}

For the empirical experiments, we use three empirical email networks with time stamps provided for all edges and test on their subnetworks. The specific procedure for these experiments is described in Sec.~\ref{subsec:methods_exp_empirical}. The first network is the ia-radoslaw-email network \cite{radoslaw}. The entire dataset is email network activity over the course of $6$ months among $167$ employee email addresses at a mid-sized manufacturing company. The second network is the Enron network as part of the Koblenz Network Collection \cite{enron}. The original network consists of more than $80\,000$ users and $1$ million emails between Enron employees from 1999 to 2003. The last network is the email-Eu-core-temporal network \cite{EUemail} generated from email data among $986$ members of a large European research institution between 2003 and 2005.

To select appropriate subnetworks, we look at all subnetworks induced by subsets of nodes $\{1,2,\ldots, n \}$ for every $1 \le n \le N$ in the entire graph $G = (V, E)$, where $|V| = N$, and the nodes' $id$s are assigned according to their first appearance in time. We then extract the subnetwork that has a significant increase in density in time, so the demonstration can be comparable to the synthetic results for how many multiples of edges are added by the end. The number of nodes represented in the following network examples refers to the subnetworks on the email users who show up first in time. Specifically, the ia-radoslaw-email subnetwork has the first $74$ nodes in time with their corresponding $1457$ edges, the Enron subnetwork is obtained by first trimming down to a 1999--2002 time frame and then selecting the first $120$ nodes in time with their associated $1603$ edges, and the email-Eu-core-temporal subnetwork is the first $282$ nodes with $4544$ edges.  

\begin{figure}[!htbp]
\includegraphics[width=8.6cm]{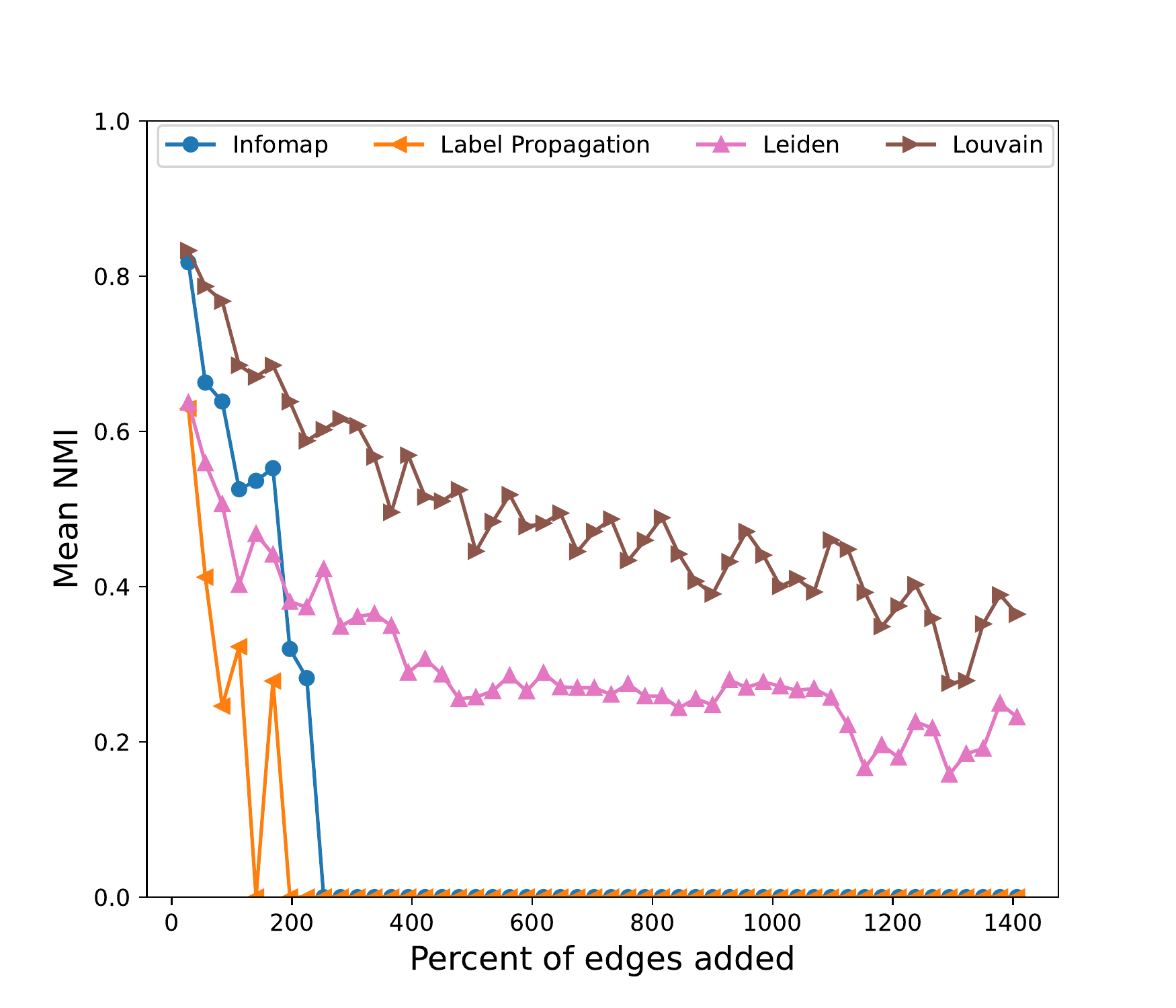}
\caption{\label{fig:rado} Mean NMI over the percentage of edges added for the ia-radoslaw-email subnetwork with 74 nodes. We use fast consensus to obtain 20 initial community partitions. Edges are added over 50 steps following time stamps from the dataset. The average NMI is computed over all pairs of the initial consensus partitions and the partitions from 10 independent realizations at each time step.}
\end{figure}

\begin{figure}[!htbp]
\includegraphics[width=8.6cm]{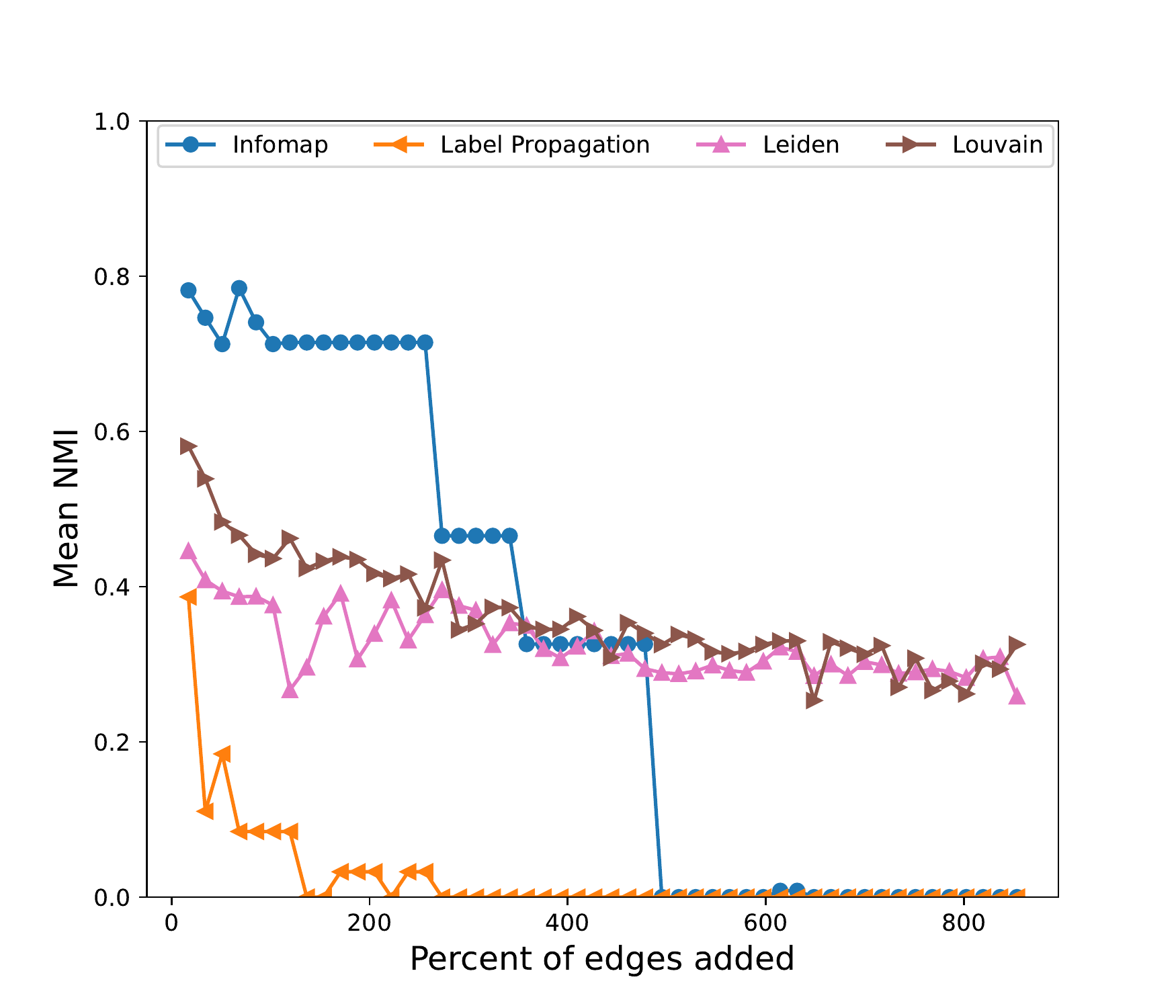}
\caption{\label{fig:enron}  Mean NMI over the percentage of edges added for the Enron subnetwork with 120 nodes.}
\end{figure}

\begin{figure}[!htbp]
\includegraphics[width=8.6cm]{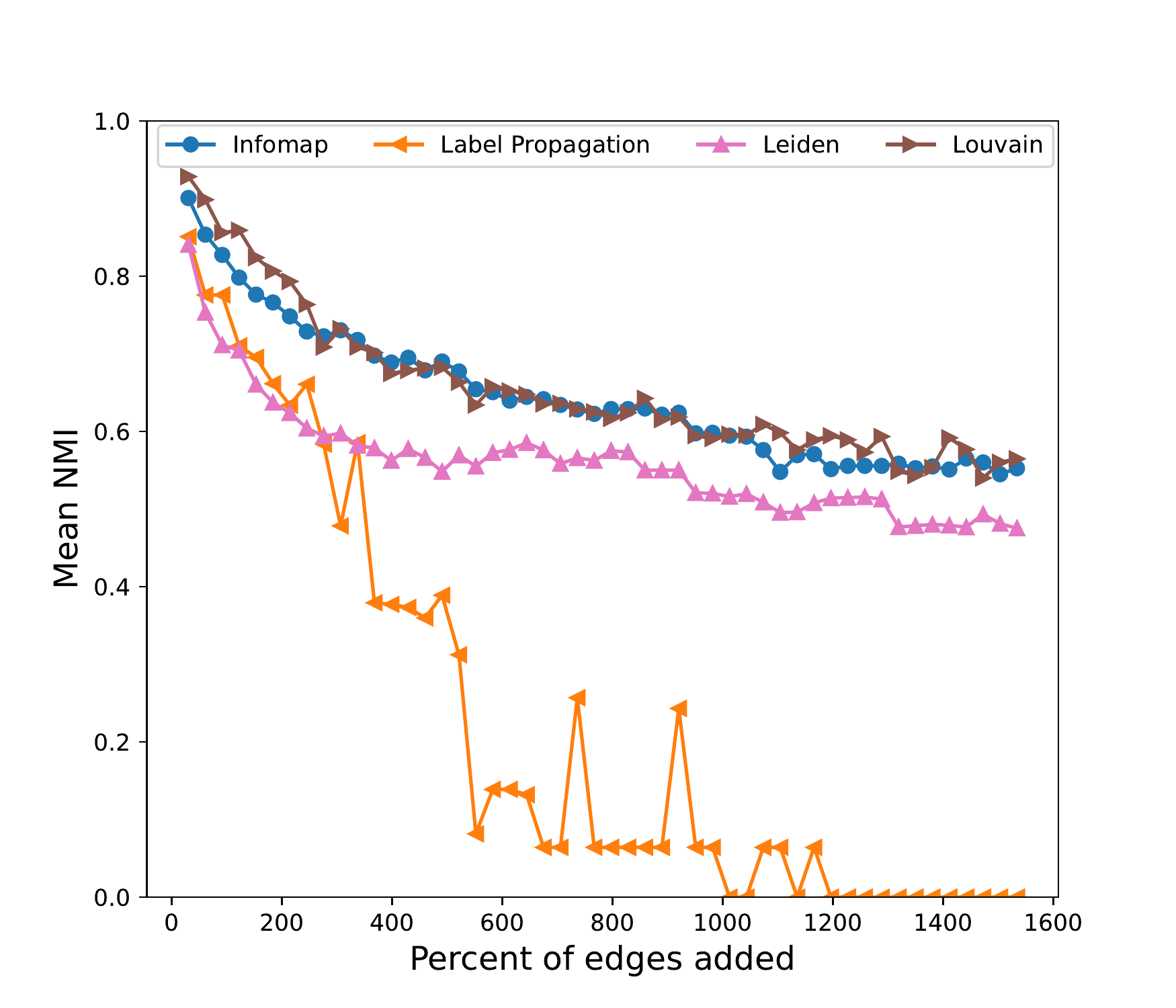}
\caption{\label{fig:eu} Mean NMI over the percentage of edges added for the email-Eu-core-temporal subnetwork with 282 nodes.}
\end{figure}

Figures~\ref{fig:rado}--\ref{fig:eu} show the mean NMI over the percentage of added edges in the subnetworks of the ia-radoslaw-email, Enron, and email-Eu-core-temporal networks, respectively. The corresponding results for standard deviation are included in Figs.~\ref{fig:rado_nmisd}--\ref{fig:eu_nmisd} in Appendix~\ref{app:emp-std}. 

Notably, empirical networks are more complex to analyze because they lack ground-truth community structure, the degree distribution might not belong to a certain family of distributions, and the set of edges added in time may be governed by different unknown factors that we cannot control.

In considering the effect of edge-addition rules in the empirical compared with the synthetic cases, we look at the ratio of intercommunity edges among all added edges over time for each empirical network. For ia-radoslaw-email, Enron, and email-Eu-core-temporal subnetworks, the intercommunity ratios are $94\%, 90\%$, and $59\%$, respectively. Note that the LFR benchmark graphs are sparse and so there is a lower fraction of intracommunity edges available to be added. Specifically, for the LFR benchmark on $1000$ nodes, the number of intracommunity nonexistent edges is only about twice the original number of edges but the number of intercommunity nonexistent edges can go to about $37\times$. This means that for our random addition on LFR benchmark with $1000$ nodes, the fraction of intercommunity edges added is around $95\%$, and due to the limited number of intracommunity edges, it is impossible to have intracommunity edges taken more than $20\%$ of the $10\times$ additional edges. The ia-radoslaw-email subnetwork has the ratio of added intercommunity edges most comparable with our previous synthetic experiments on $1000$ nodes. To draw a direct comparison between the empirical and the synthetic cases, we experiment on synthetic networks where the ratio of added intercommunity edges is controlled to match the one in ia-radoslaw-email subnetwork, namely $94\%$. The results show similar behaviors as the previous synthetic ones and we include them in Appendix~\ref{app:syn-matchRatio}.

In addition, we acknowledge that the chosen subnetworks described here are not as large as the synthetic benchmark graphs: The benchmark graphs have thousands of nodes, whereas the empirical network examples only have hundreds.

Among these results in NMI, the experiments on the ia-radoslaw-email and Enron subnetworks reveal similar behaviors as the synthetic networks (i.e., Leiden and Louvain appear to detect more sustainable community structure than Infomap or Label Propagation as more edges are added). However, in the experiments on the email-Eu-core-temporal subnetwork, Infomap has performance similar to Leiden and Louvain, whereas Label Propagation has the lowest NMI score almost throughout all time steps. Although there may be intertwining reasons for these results, one shared phenomenon we observe for all subnetworks is that with Infomap or Label Propagation, the mean NMI drops to almost $0$, whenever the detection algorithm begins to detect only one community in the perturbed networks. 

\begin{figure}[!htbp]
\includegraphics[width=8.6cm]{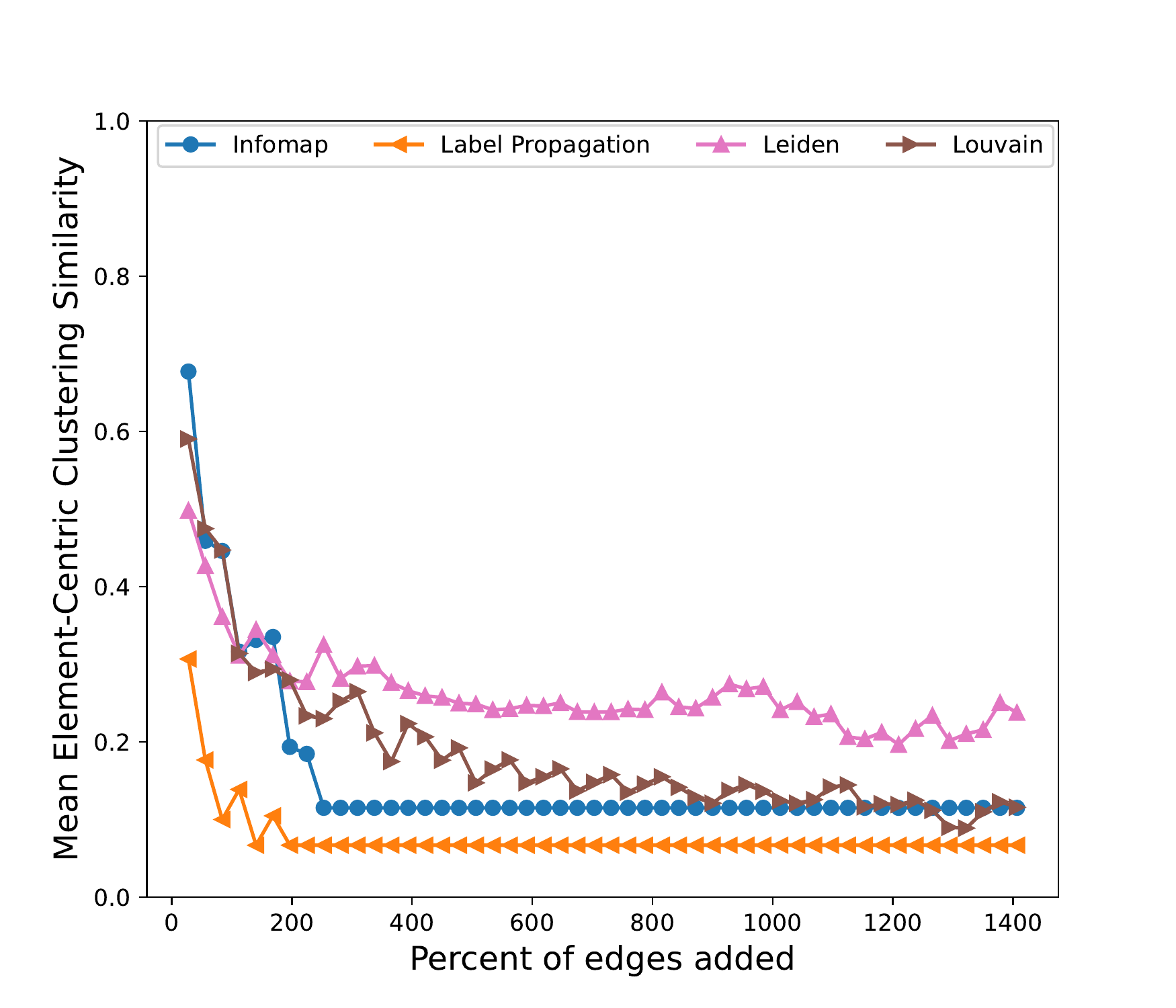}
\caption{\label{fig:rado_elsim} Mean element-centric clustering similarity over the percentage of edges added for the ia-radoslaw-email subnetwork.}
\end{figure}

\begin{figure}[!htbp]
\includegraphics[width=8.6cm]{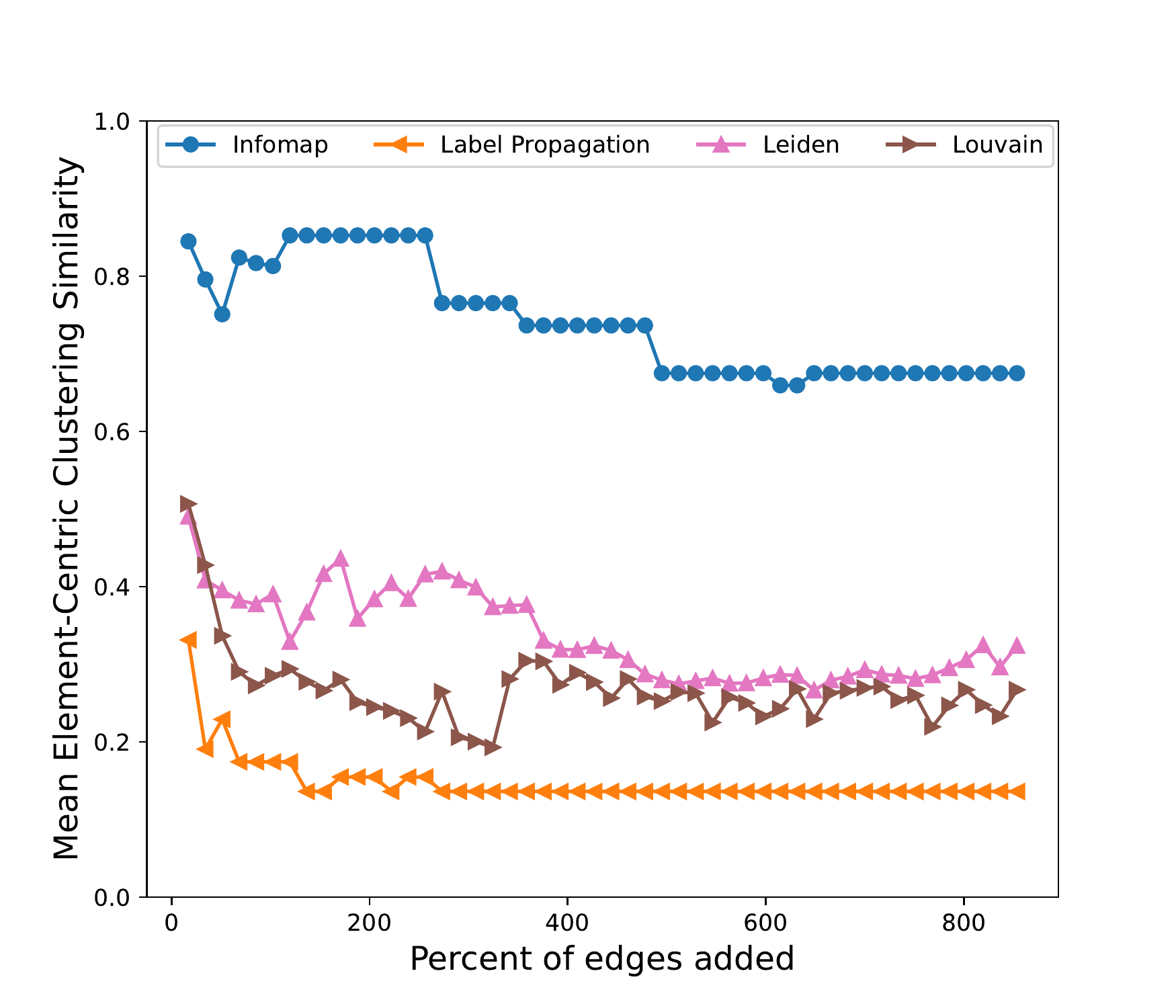}
\caption{\label{fig:enron_elsim} Mean element-centric clustering similarity over the percentage of edges added for the Enron subnetwork.}
\end{figure}

\begin{figure}[!htbp]
\includegraphics[width=8.6cm]{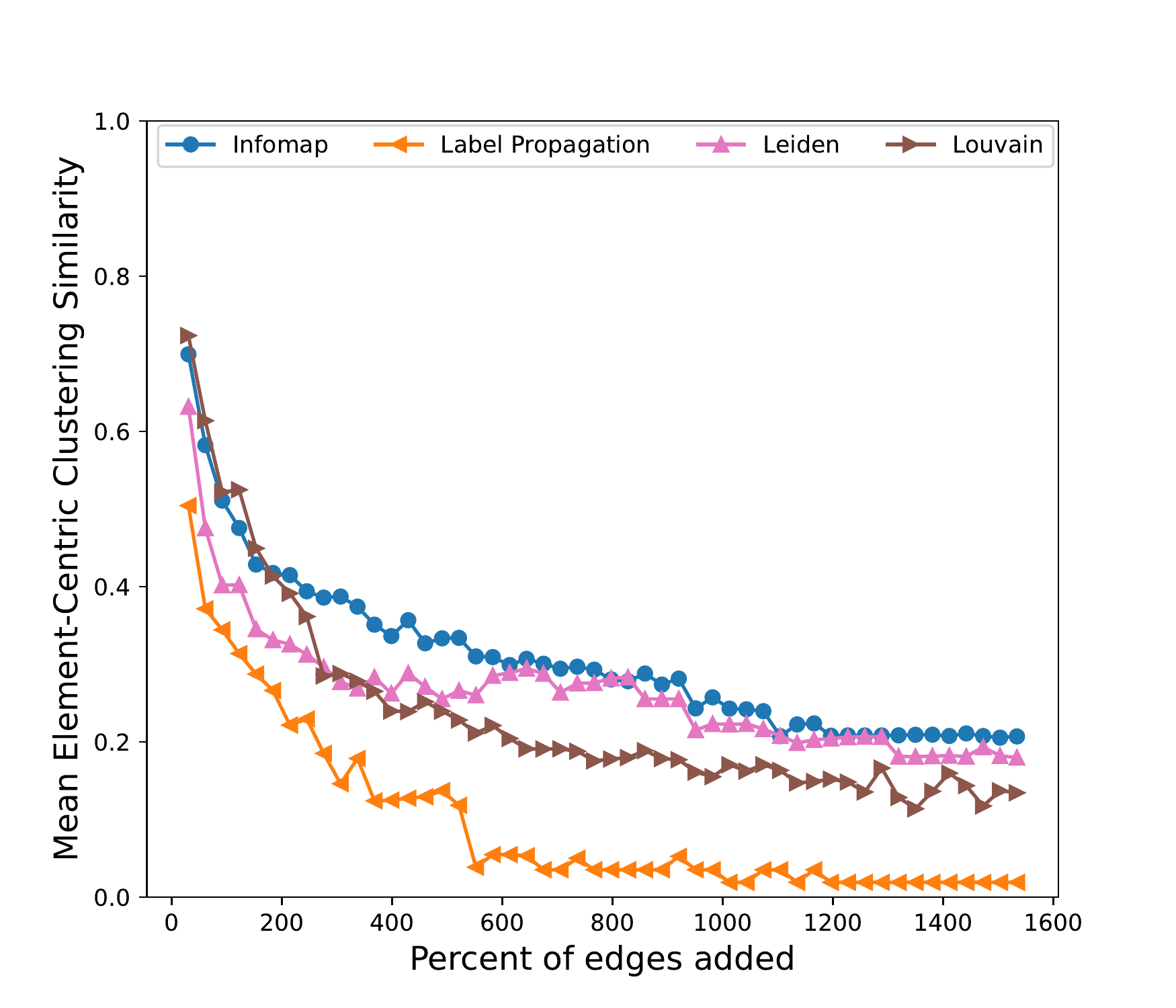}
\caption{\label{fig:eu_elsim} Mean element-centric clustering similarity over the percentage of edges added for the email-Eu-core-temporal subnetwork.}
\end{figure}

While community detection algorithms have different performance, we also observe effects from using different community similarity metrics. Specifically, we repeat the same sets of experiments but replace the NMI metric with the element-centric clustering similarity. Figures~\ref{fig:rado_elsim}--\ref{fig:eu_elsim} illustrate these results. The corresponding standard deviations are provided in Figs.~\ref{fig:rado_elsimsd}--\ref{fig:eu_elsimsd} in Appendix~\ref{app:emp-std}. Using this different metric, we find that Infomap, Leiden, and Louvain do not show consistent advantages over each other, but Label Propagation, although not necessarily dropping to $0$ by the end of time, always has the lowest metric value.

One noticeable difference in these element-centric clustering similarity results from the NMI results is the curve for Infomap in the Enron subnetwork. In the NMI plot (Fig.~\ref{fig:enron}), the value eventually drops to $0$, which shows that the community structure detected by Infomap is not as robust as that by Leiden or Louvain in terms of NMI. However, in the element-centric clustering similarity metric, the clustering similarity value is the highest for Infomap throughout time among the four algorithms, and this suggests that the community structure found by Infomap is the most robust in our experiments according to this alternative metric. To determine what is happening, we look at the mean number of communities detected by the four algorithms at each step and find that it drops from higher numbers to $1$ for both Infomap and Label Propagation but maintains around $5$ or $6$ for Leiden and Louvain by the end of time. Note that the NMI always outputs $0$ whenever one of the two partitions being compared has only one community due to the formulation of this metric, so we suspect that this property might have hindered NMI from capturing some similarities between the communities in the initial and the highly perturbed networks, especially after the number of detected communities drops to $1$.

Nonetheless, when networks are perturbed under edge addition, and the density is increased by a significant amount, different community detection algorithms start to show clear discrepancies in whether they can find a community structure similar to the initial one. Hence, the chosen community detection algorithm plays an important role in detecting robust community structures over time.

\section{\label{sec:level5} Conclusion}

We design synthetic and empirical experiments to test the robustness of community structure under the perturbation of edge addition by using different community detection algorithms. Overall, we found that community robustness strongly depends on the community detection algorithm selected. In the synthetic experiments, we use LFR benchmark graphs and control the mixing parameter, $\mu$. To mimic how edges may be added in different scenarios in real networks and to illustrate the difference in the outputs, we add random edges in two different ways. Both ways select additional edges from nonexistent edges. One is a completely uniformly random selection, which is analogous to random errors, and the other is a targeted selection from edges across different communities, which is analogous to attacks in real-world networks. 

We demonstrate results for six different mixing parameter values and the two different edge-addition methods described above. The clustering similarity scores computed in NMI indicate that networks with lower mixing parameters (i.e. stronger partitions) have more robust community structures under the perturbation of edge addition. Our targeted edge-addition method can more efficiently alter the initial communities compared with the random addition. As expected, the NMI values drop faster in the targeted case among all chosen community detection algorithms. 

In these synthetic experiments, modularity-based algorithms, Leiden and Louvain, show better performance in detecting more similar communities to the initial ones vs Infomap and Label Propagation, which cannot detect any community structure when graphs become too dense. In other words, Leiden and Louvain excel at finding more robust communities in networks that can withstand more severe network perturbations. We also observe effects caused by community similarity metrics, NMI and element-centric clustering similarity specifically, but the takeaways in the two metrics are not significantly different in the qualitative sense. The overall impacts of the initial partition strength, the edge-addition method, and the selected community detection algorithm are similar with either metric.

We acknowledge that empirical temporal networks introduce more complexity, and we see different community similarity metrics demonstrate significantly different results when determining whether the detected communities are similar to the initial ones (i.e., whether the community detection algorithm can find a robust community structure). In the empirical experiments, we again find that community detection algorithms play an important role in the robustness of communities. Specifically, for all three empirical subnetworks tested, we observe that Label Propagation performs the worst among the four algorithms in detecting robust communities over time using either NMI or element-centric clustering similarity. 

When considering future research directions, we note that the different metrics and community detection algorithms show different outcomes on the empirical temporal networks. To understand more about their effects on the community robustness performance, we need many more network examples in which edge densities expand over time. Adequate network candidates with properties in different families---including scales, densities, degree distributions, and edge-addition rules---are needed for comparison tests in order to distinguish the effects from each individual aspect. Another direction for future research is to explore different types of generative models for benchmarking and include community detection algorithms based on other methods for comparison.

\begin{acknowledgments}
This paper has been authored by UT-Battelle, LLC under Contract No. DE-AC05-00OR22725 with the U.S. Department of Energy. The publisher, by accepting the article for publication, acknowledges that the U.S. government retains a nonexclusive, paid up, irrevocable, world-wide license to publish or reproduce the published form of the manuscript, or allow others to do so, for U.S. government purposes. The DOE will provide public access to these results in accordance with the DOE Public Access Plan (http://energy.gov/downloads/doe-public-access-plan).

This research was sponsored in part by Oak Ridge National Laboratory's (ORNL's) Laboratory Directed Research and Development program and by the U.S. Department of Energy. M.T. acknowledges support from the National Science Foundation Mathematical Sciences Graduate Internship program. The funders had no role in study design, data collection and analysis, decision to publish, or preparation of the manuscript. We also thank Bj\"{o}rn Sandstede for providing critical feedback on the manuscript and Matthew T. Harrison and Ramakrishnan Kannan for providing inspirational suggestions to the project over our conversations.
\end{acknowledgments}

\clearpage
\appendix
\section{\label{app:syn-clusim}Synthetic Results Using Element-Centric Clustering Similarity Metric}

\begin{figure}[b]
\includegraphics[width=8.6cm]{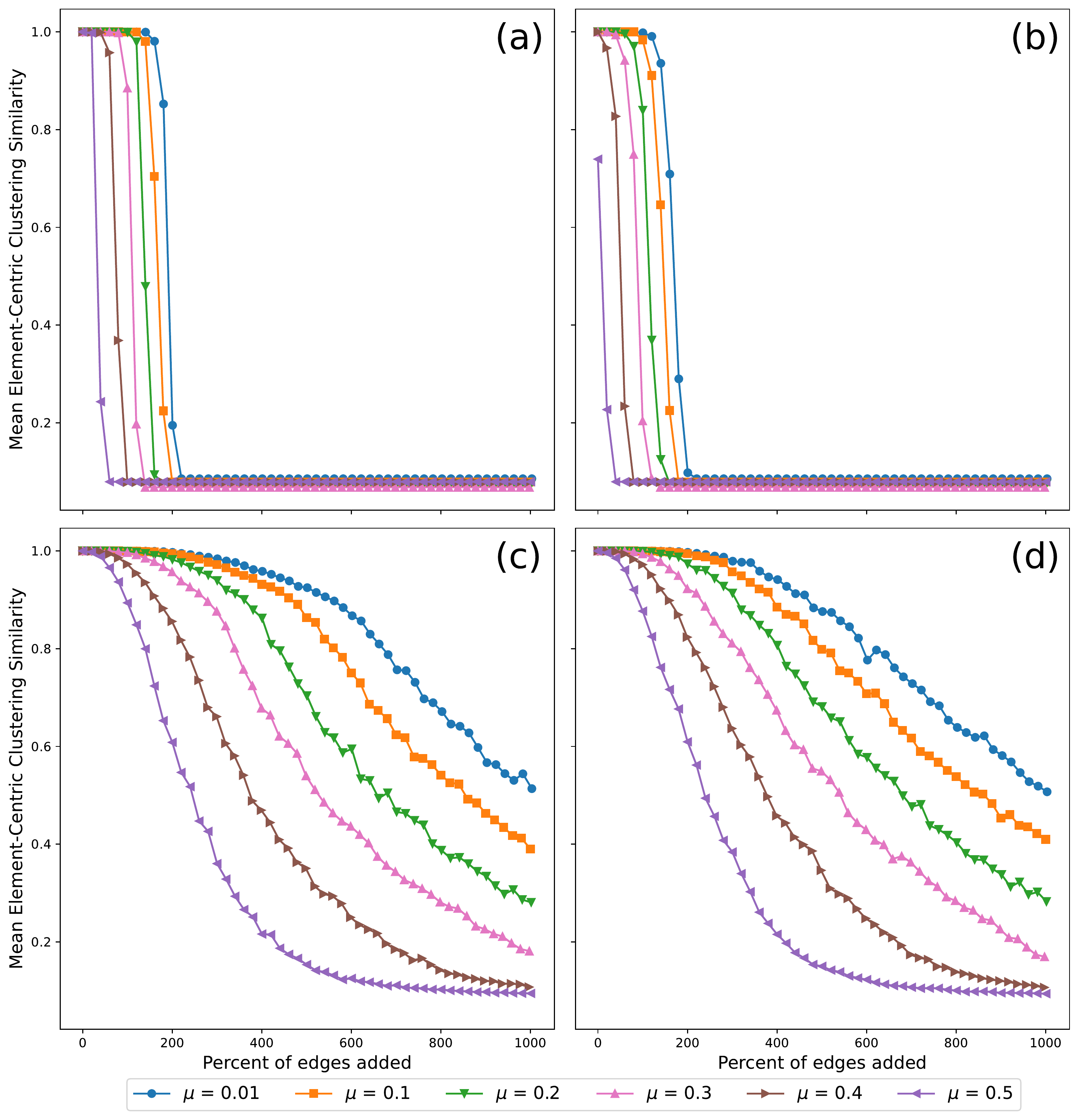}
\caption{\label{fig:LFR1000_elsim} Mean element-centric clustering similarity over the percentage of edges added uniformly at random on LFR benchmark graphs with $1000$ nodes. Communities detected by (a)~Infomap, (b)~Label Propagation, (c)~Leiden, and (d)~Louvain.}
\end{figure}

\begin{figure}[!ht]
\includegraphics[width=8.6cm]{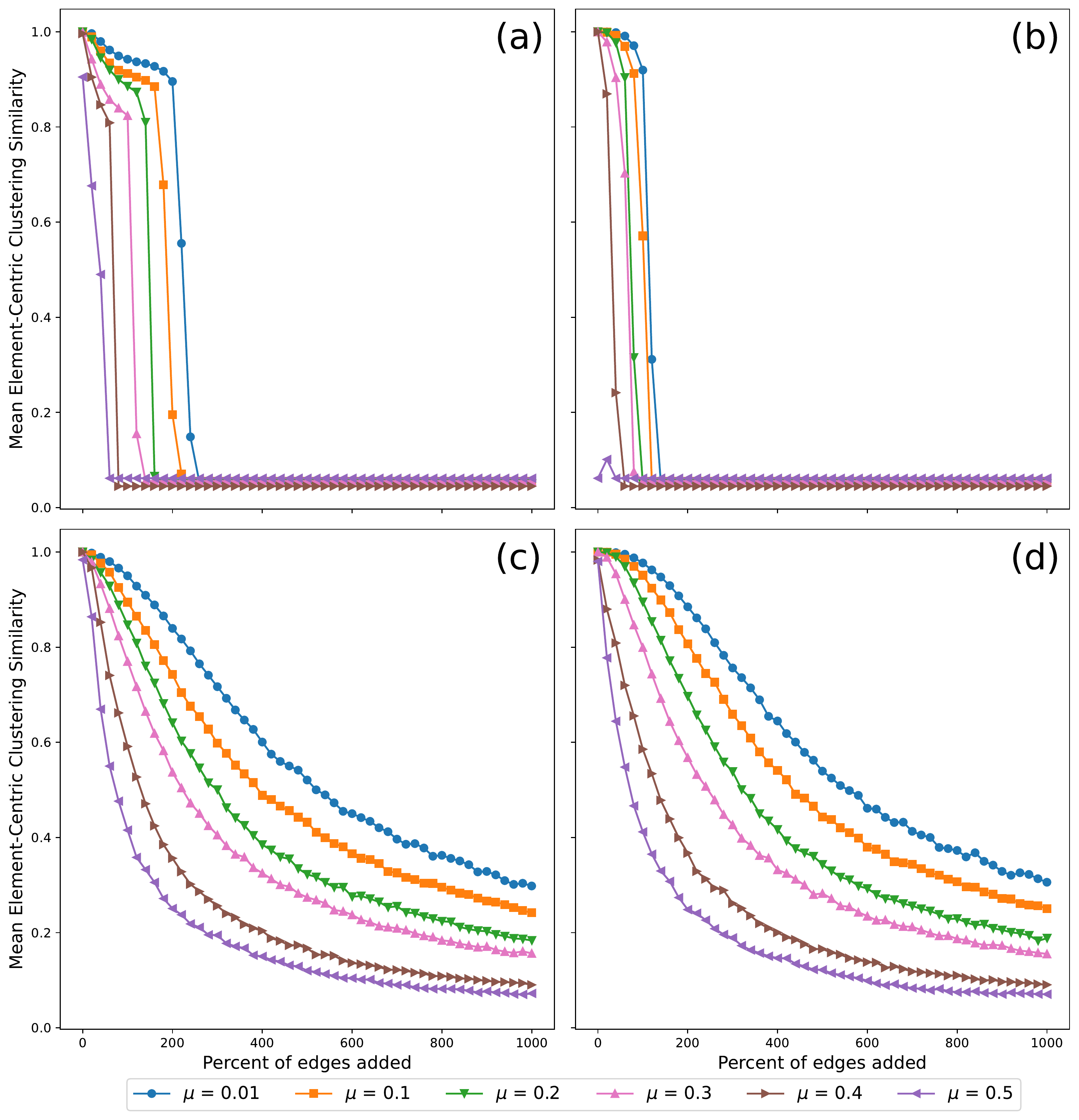}
\caption{\label{fig:LFR10000_elsim} Mean element-centric clustering similarity over the percentage of edges added uniformly at random on LFR benchmark graphs with $10\,000$ nodes. Communities detected by (a)~Infomap, (b)~Label Propagation, (c)~Leiden, and (d)~Louvain.}
\end{figure}

\begin{figure}[H]
\includegraphics[width=8.6cm]{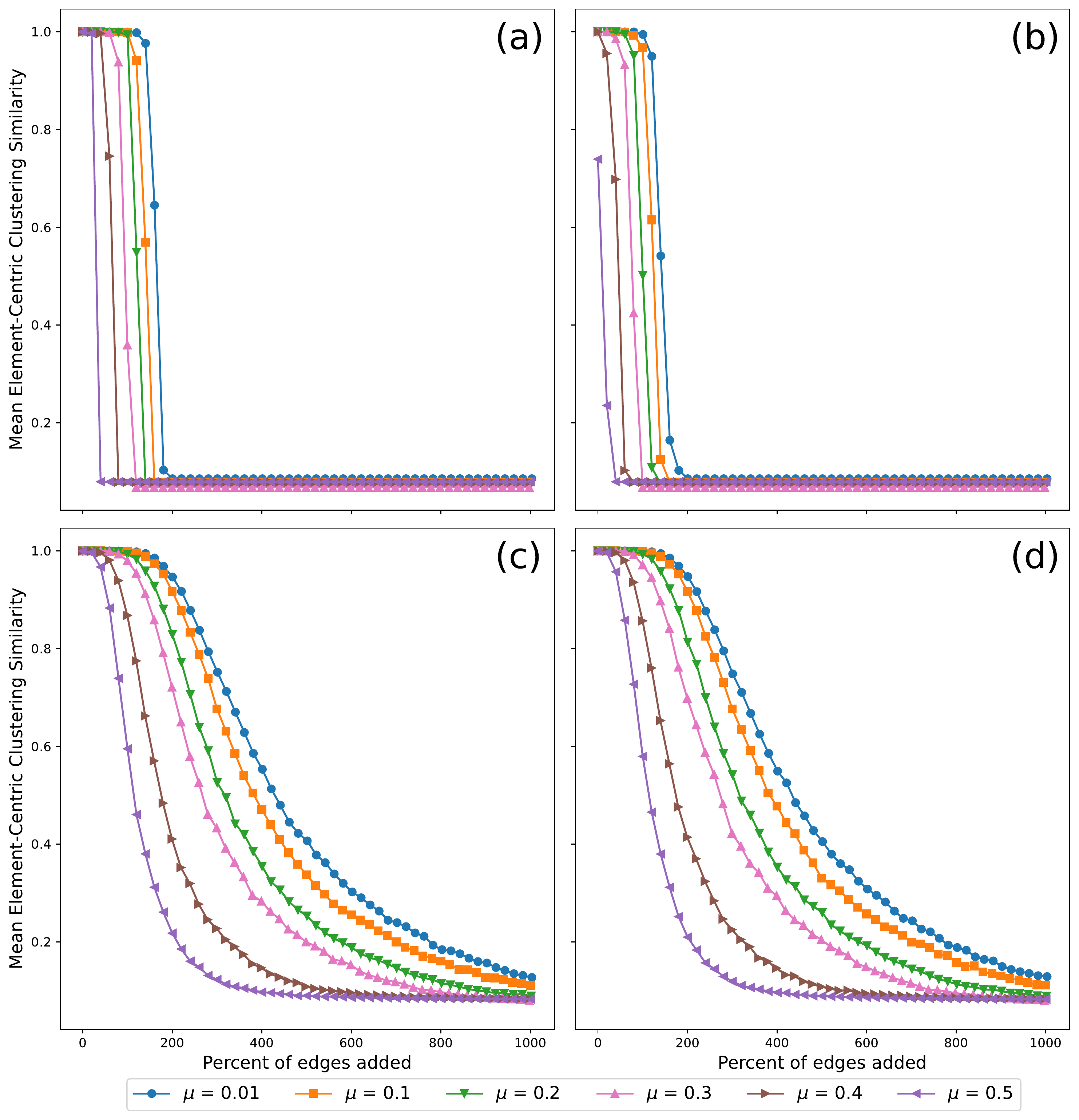}
\caption{\label{fig:LFR1000_elsim_target} Mean element-centric clustering similarity over the percentage of edges added that are selected uniformly at random across different communities on LFR benchmark graphs with $1000$ nodes. Communities detected by (a)~Infomap, (b)~Label Propagation, (c)~Leiden, and (d)~Louvain.}
\end{figure}

\begin{figure}[!ht]
\includegraphics[width=8.6cm]{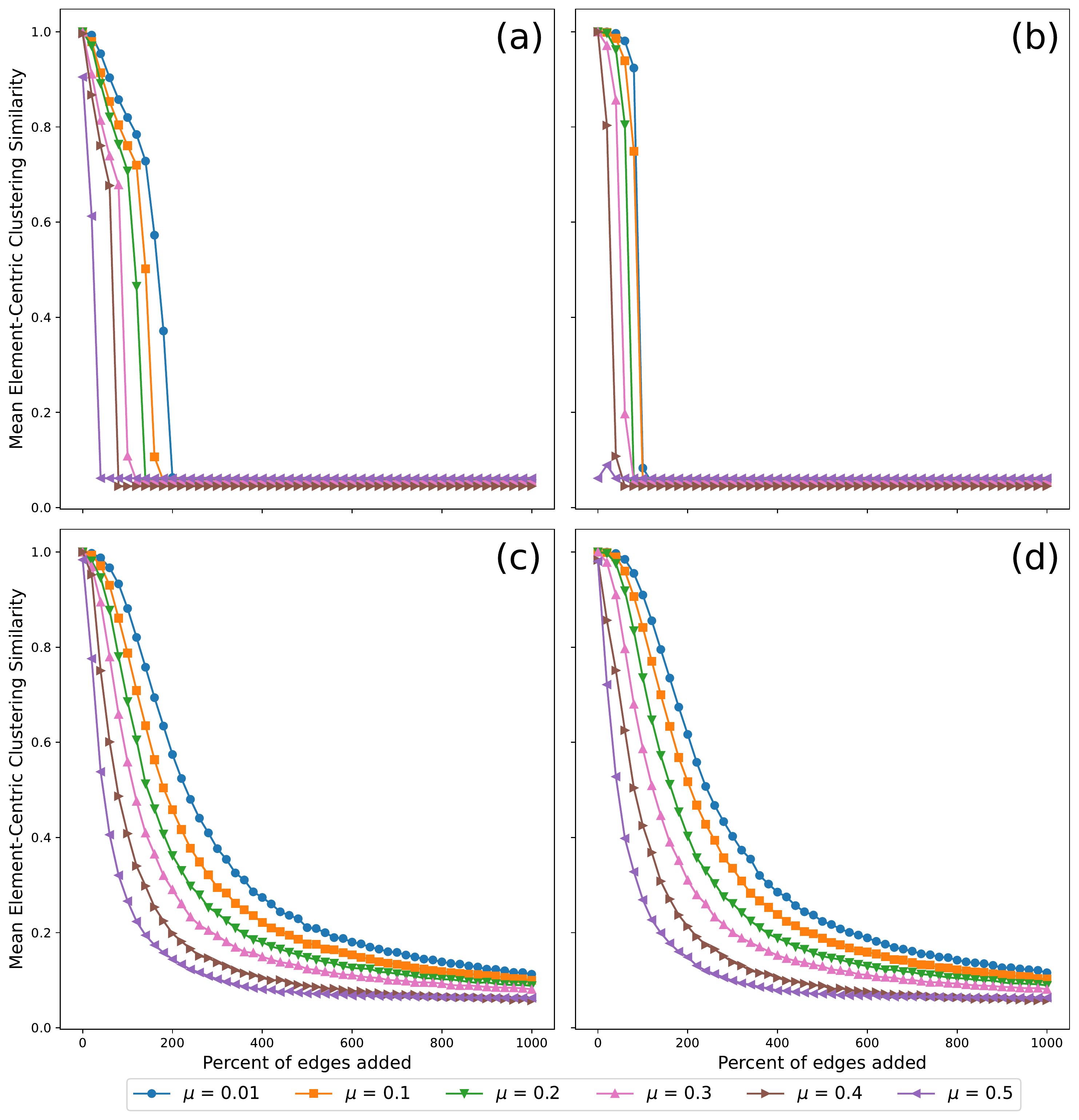}
\caption{\label{fig:LFR10000_elsim_target} Mean element-centric clustering similarity over the percentage of edges added that are selected uniformly at random across different communities on LFR benchmark graphs with $10\,000$ nodes. Communities detected by (a)~Infomap, (b)~Label Propagation, (c)~Leiden, and (d)~Louvain.}
\end{figure}

Figures~\ref{fig:LFR1000_elsim} and~\ref{fig:LFR10000_elsim} show the results of using the element-centric clustering similarity metric for the LFR benchmark graphs with $1000$ nodes and $10\,000$ nodes, respectively, with added edges selected uniformly at random from all nonexistent edges while multiedges are prohibited. Notably, the parameter values are chosen the same as in experiments that use NMI, and the experimental procedure follows Sec.~\ref{subsec:methods_exp_synthetic}.
Figures~\ref{fig:LFR1000_elsim_target} and~\ref{fig:LFR10000_elsim_target} show the LFR benchmark graphs with $1000$ nodes and $10\,000$ nodes, respectively, with additional edges selected uniformly at random but restricted to ones that cross different communities.

Results in element-centric clustering similarity agree with the observations from results in NMI. Networks with stronger initial partitions, specifically those with lower $\mu$ values, are relatively more robust in the community structure. Targeted edge addition can more quickly destroy the original community structure. Also, community robustness is highly dependent on the chosen community detection algorithms. In particular, Leiden and Louvain can detect communities similar to the initial ground truth as the graphs increase the number of edges significantly.

\newpage

\section{\label{app:syn-matchRatio} Synthetic Results Matching Ratio of Intercommunity Edges in Empirical Experiments}

\begin{figure}[!ht]
\includegraphics[width=8.6cm]{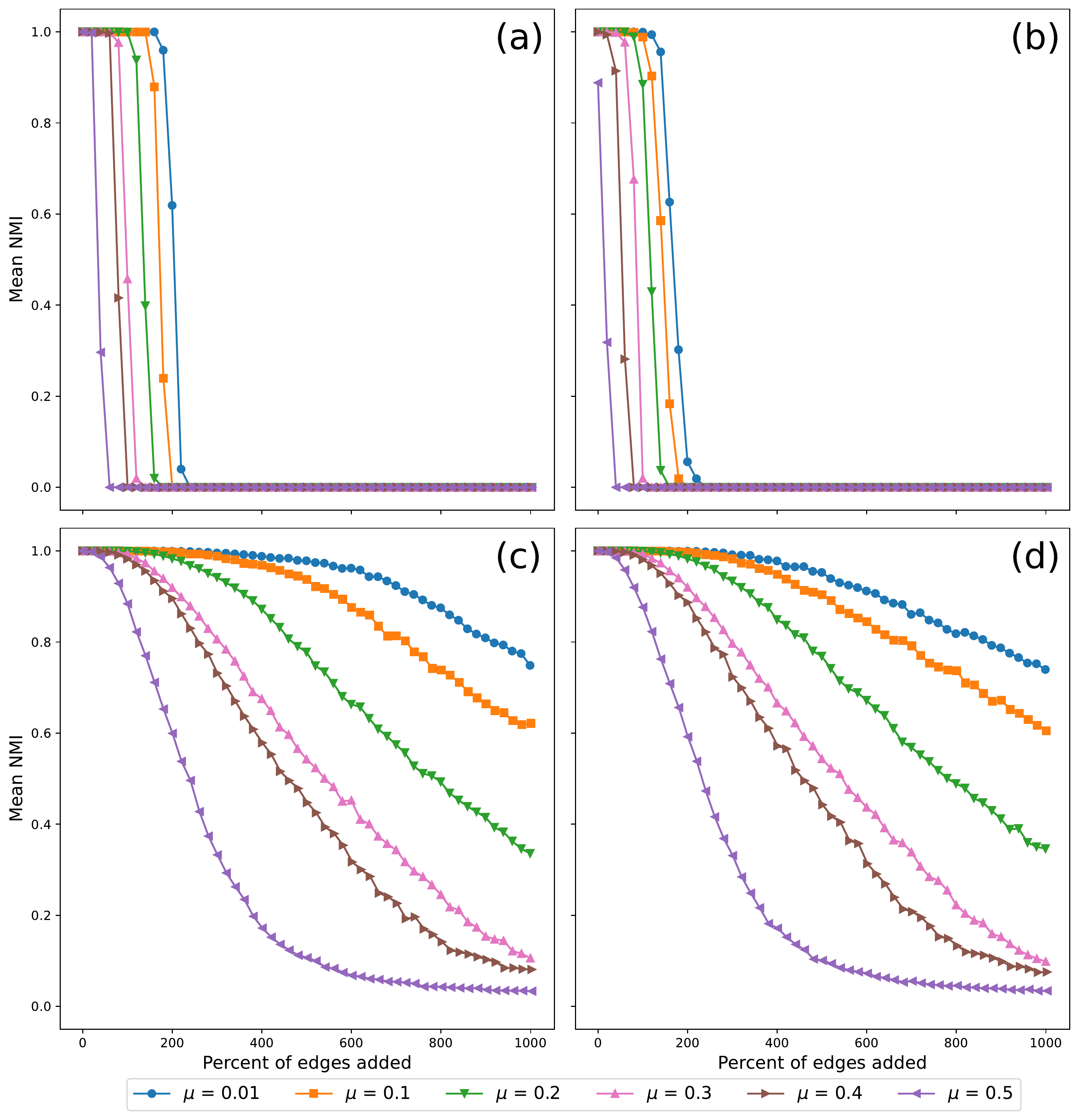}
\caption{\label{fig:LFR1000_nmi_r94} Mean NMI over the percentage of edges added on LFR benchmark graphs with $1000$ nodes. The ratio of intercommunity edges added is $94\%$, which matches the one in the ia-radoslaw-email subnetwork. Communities detected by (a)~Infomap, (b)~Label Propagation, (c)~Leiden, and (d)~Louvain.}
\end{figure}

\begin{figure}[!ht]
\includegraphics[width=8.6cm]{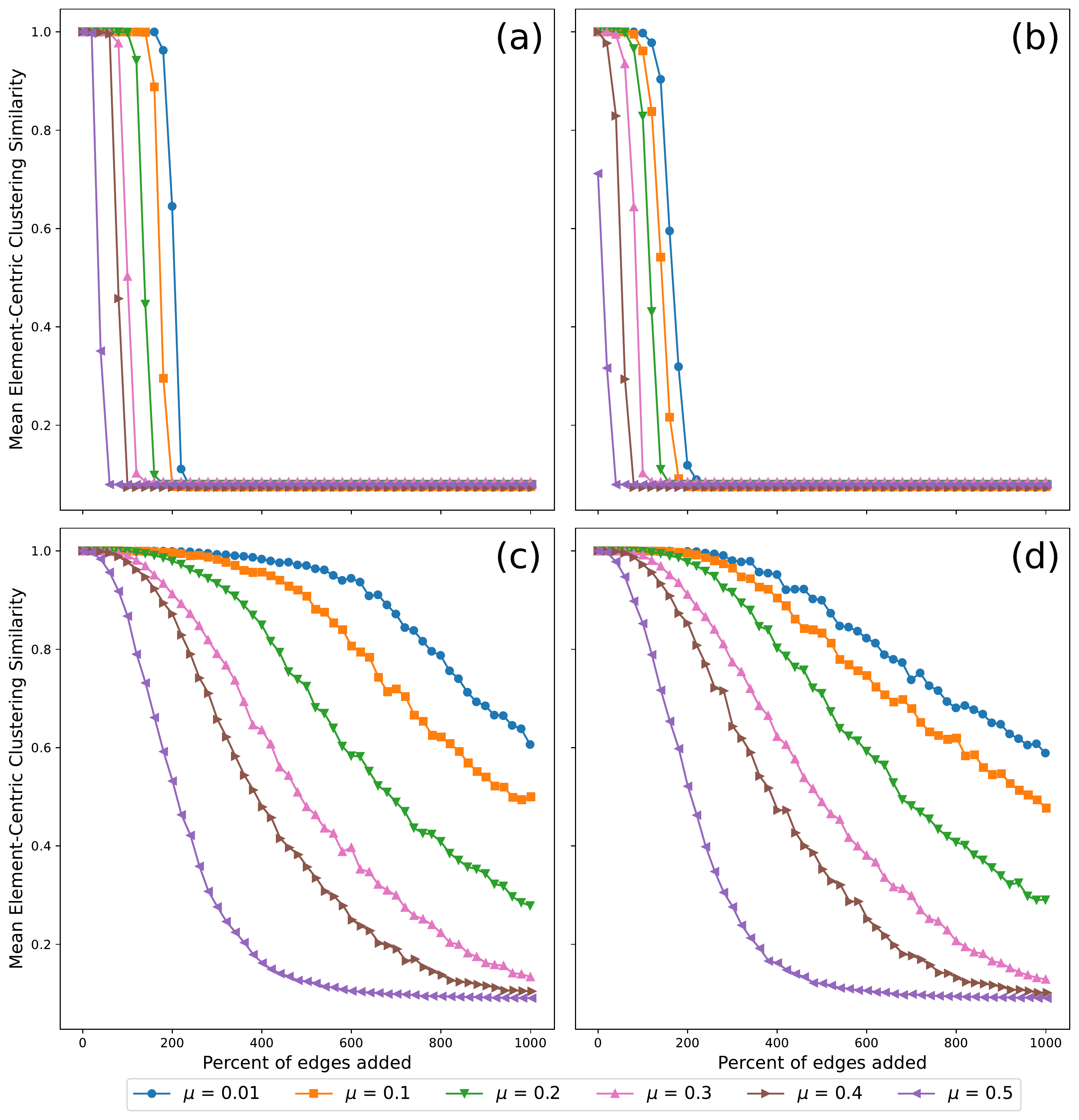}
\caption{\label{fig:LFR10000_elsim_r94} Mean element-centric clustering similarity over the percentage of edges added on LFR benchmark graphs with $1000$ nodes. The ratio of intercommunity edges added is $94\%$. Communities detected by (a)~Infomap, (b)~Label Propagation, (c)~Leiden, and (d)~Louvain.}
\end{figure}

Figures~\ref{fig:LFR1000_nmi_r94} and \ref{fig:LFR10000_elsim_r94} show mean community similarity metrics, NMI and element-centric clustering similarity respectively, over percentage of edges added on LFR benchmark graphs with $1000$ nodes. Here, at each step, we force $94\%$ of the added edges to be randomly selected from the pool of intercommunity nonexistent edges. This ratio matches the proportion we found for the intercommunity edges added of the total additional edges in the ia-radoslaw-email subnetwork. Specifically, since we have four community detection methods and each of them has $np=20$ initial communities found by the fast consensus algorithm, we have $80$ corresponding ratios and then we take the mean as the estimate.

These results show similar behaviors as those observed in the synthetic experiments: A lower mixing parameter, $\mu$, tends to have a more robust community structure, and among the four community detection algorithms, Leiden and Louvain can detect communities more similar to the initial ground truth.

\clearpage
\section{\label{app:syn-std} Standard Deviation for Synthetic Results}

\begin{figure}[h]
\centering
\includegraphics[width=8.6cm]{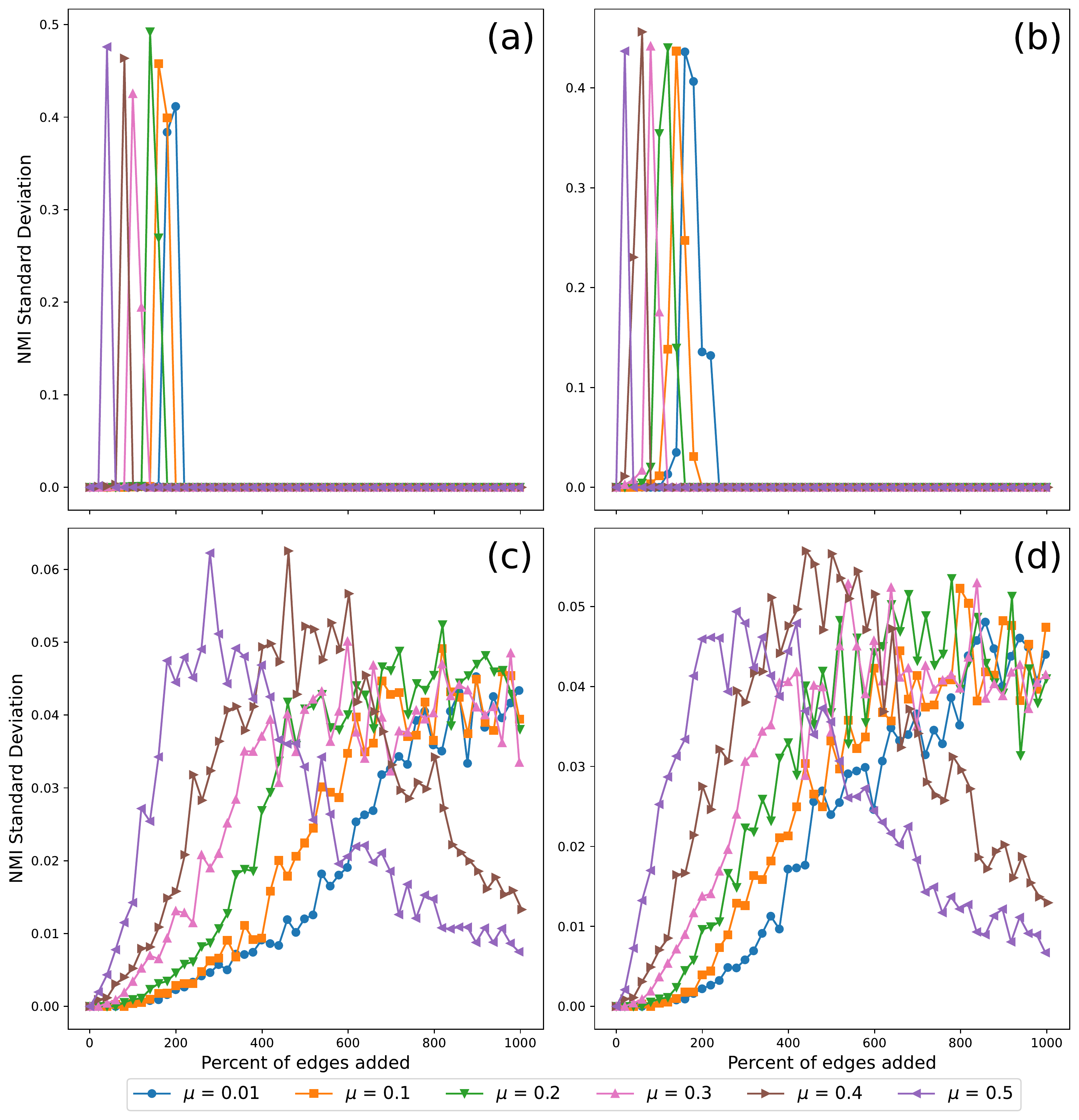}
\caption{\label{fig:LFR1000_nmisd} Standard deviation of NMI over the percentage of edges added uniformly at random on LFR benchmark graphs with $1000$ nodes. Communities detected by (a)~Infomap, (b)~Label Propagation, (c)~Leiden, and (d)~Louvain.}
\end{figure}

\begin{figure}[h]
\centering
\includegraphics[width=8.6cm]{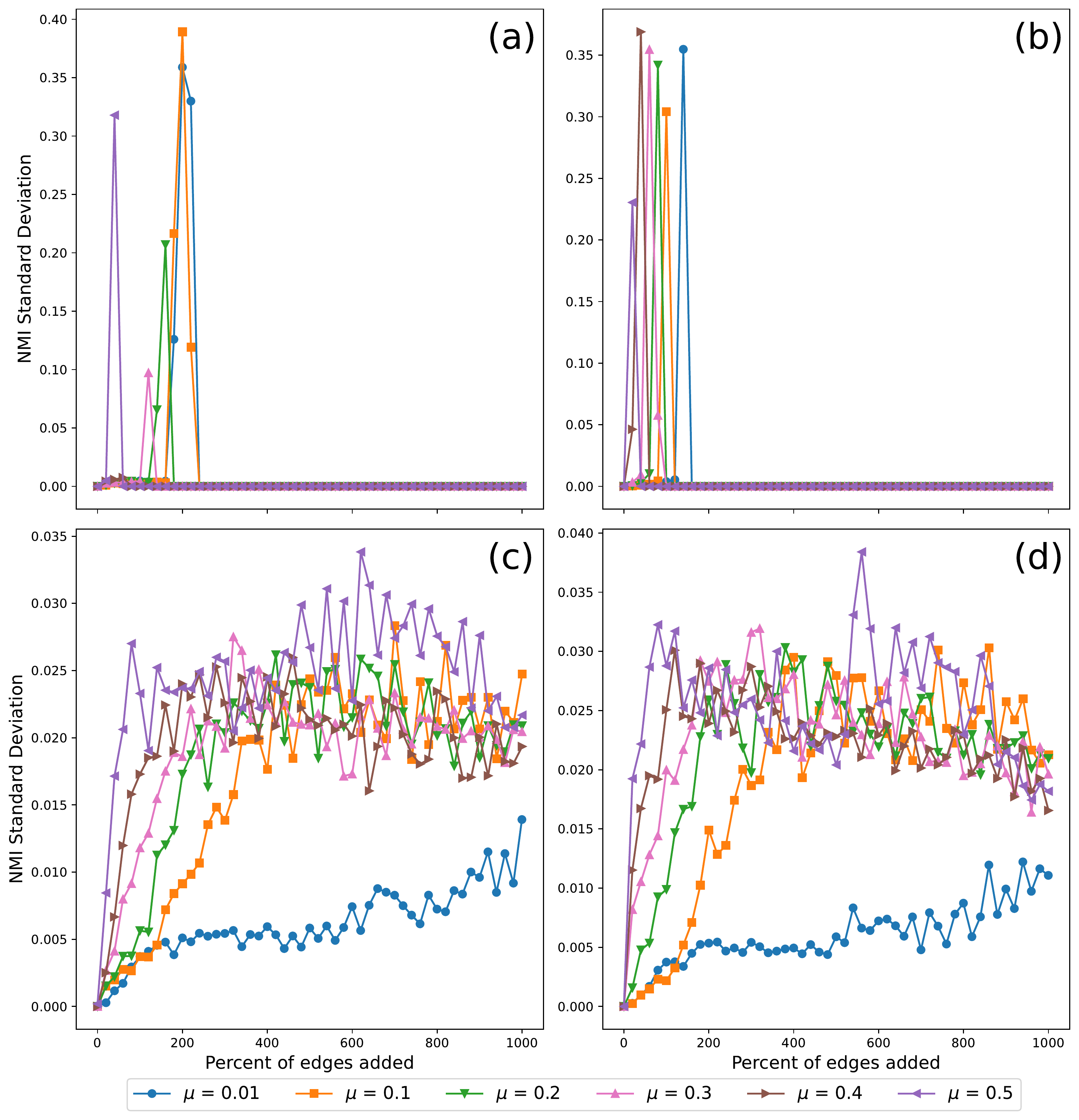}
\caption{\label{fig:LFR10000_nmisd} Standard deviation of NMI over the percentage of edges added uniformly at random on LFR benchmark graphs with $10\,000$ nodes. Communities detected by (a)~Infomap, (b)~Label Propagation, (c)~Leiden, and (d)~Louvain.}
\end{figure}

\begin{figure}[H]
\centering
\includegraphics[width=8.6cm]{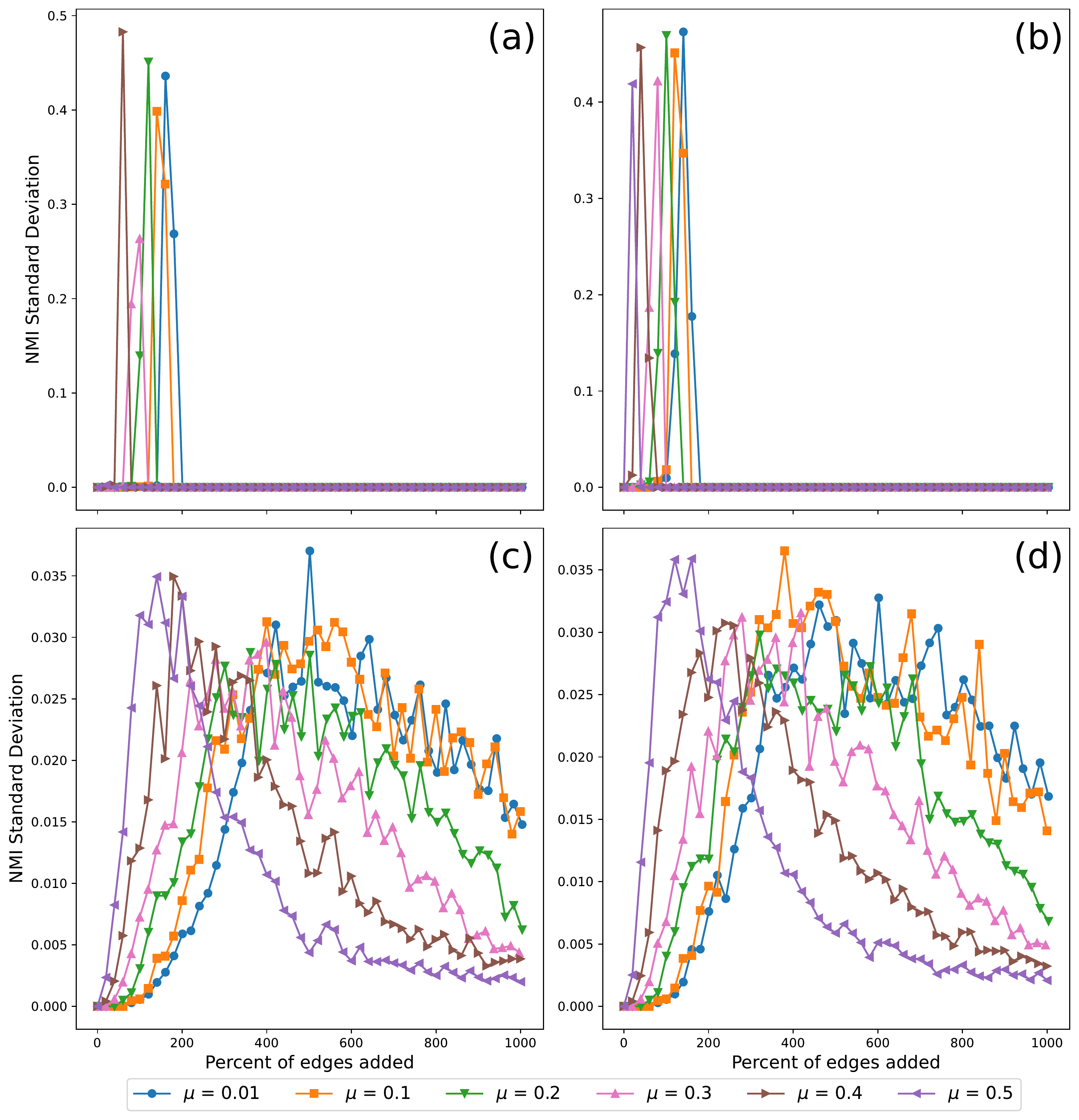}
\caption{\label{fig:LFR1000_nmisd_target} Standard deviation of NMI over the percentage of edges added that are selected uniformly at random across different communities on LFR benchmark graphs with $1000$ nodes. Communities detected by (a)~Infomap, (b)~Label Propagation, (c)~Leiden, and (d)~Louvain.}
\end{figure}

\begin{figure}[!t]
\centering
\includegraphics[width=8.6cm]{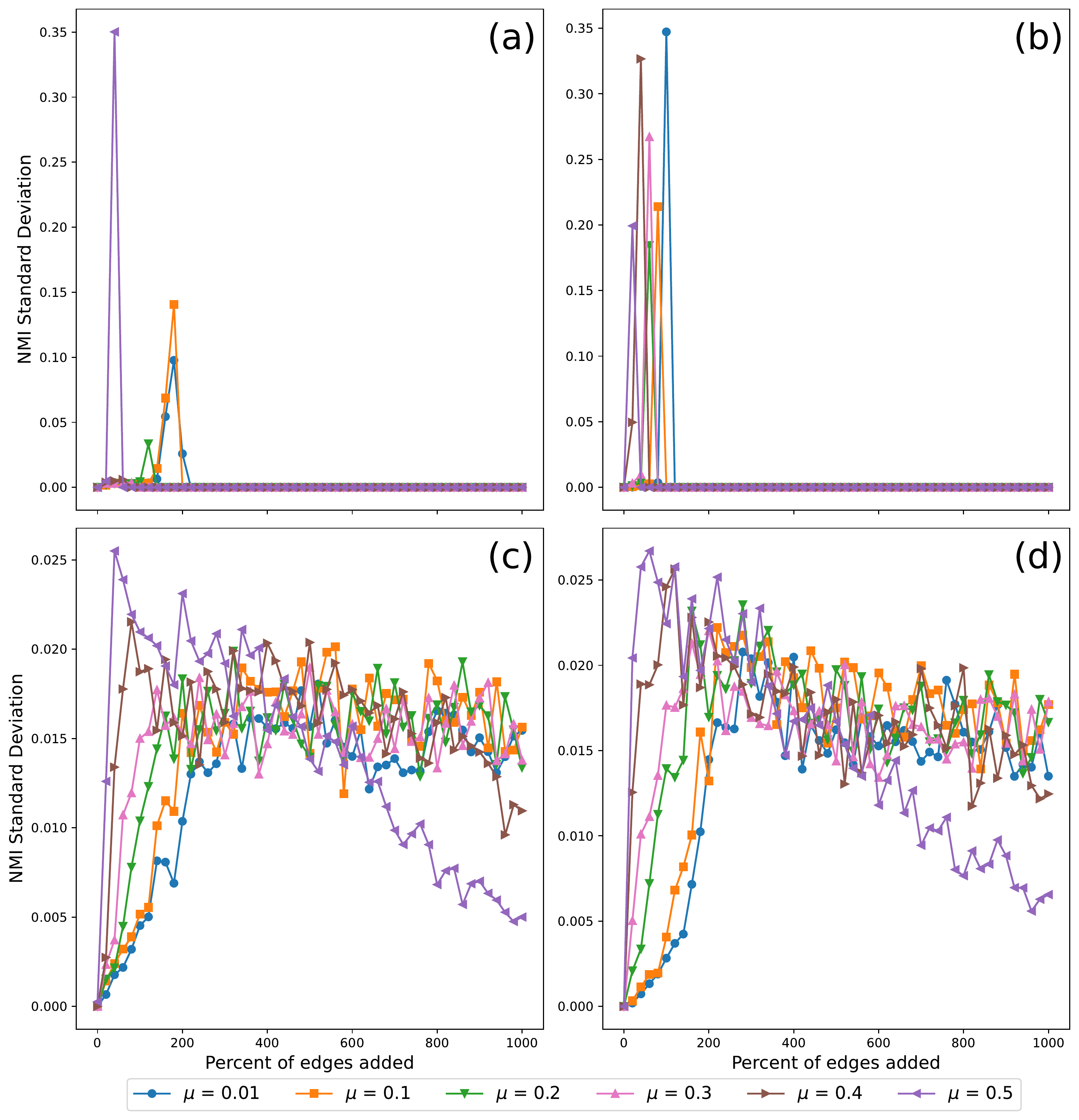}
\caption{\label{fig:LFR10000_nmisd_target} Standard deviation of NMI over the percentage of edges added that are selected uniformly at random across different communities on LFR benchmark graphs with $10\,000$ nodes. Communities detected by (a)~Infomap, (b)~Label Propagation, (c)~Leiden, and (d)~Louvain.}
\end{figure}

\begin{figure}[!h]
\centering
\includegraphics[width=8.4cm]{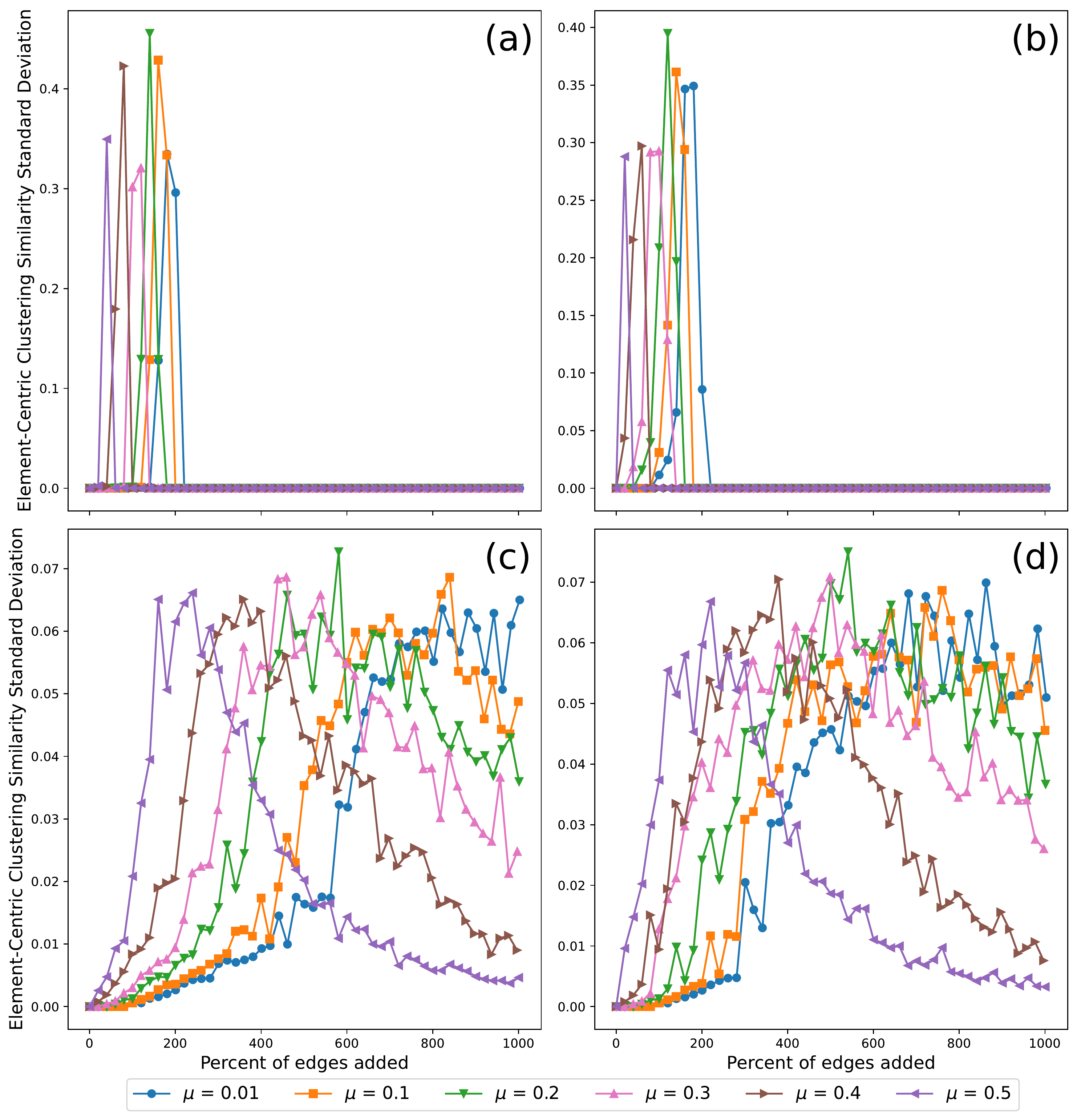}
\caption{\label{fig:LFR1000_elsimsd} Standard deviation of element-centric clustering similarity over the percentage of edges added uniformly at random on LFR benchmark graphs with $1000$ nodes. Communities detected by (a)~Infomap, (b)~Label Propagation, (c)~Leiden, and (d)~Louvain.}
\end{figure}

\begin{figure}[!t]
\centering
\includegraphics[width=8.4cm]{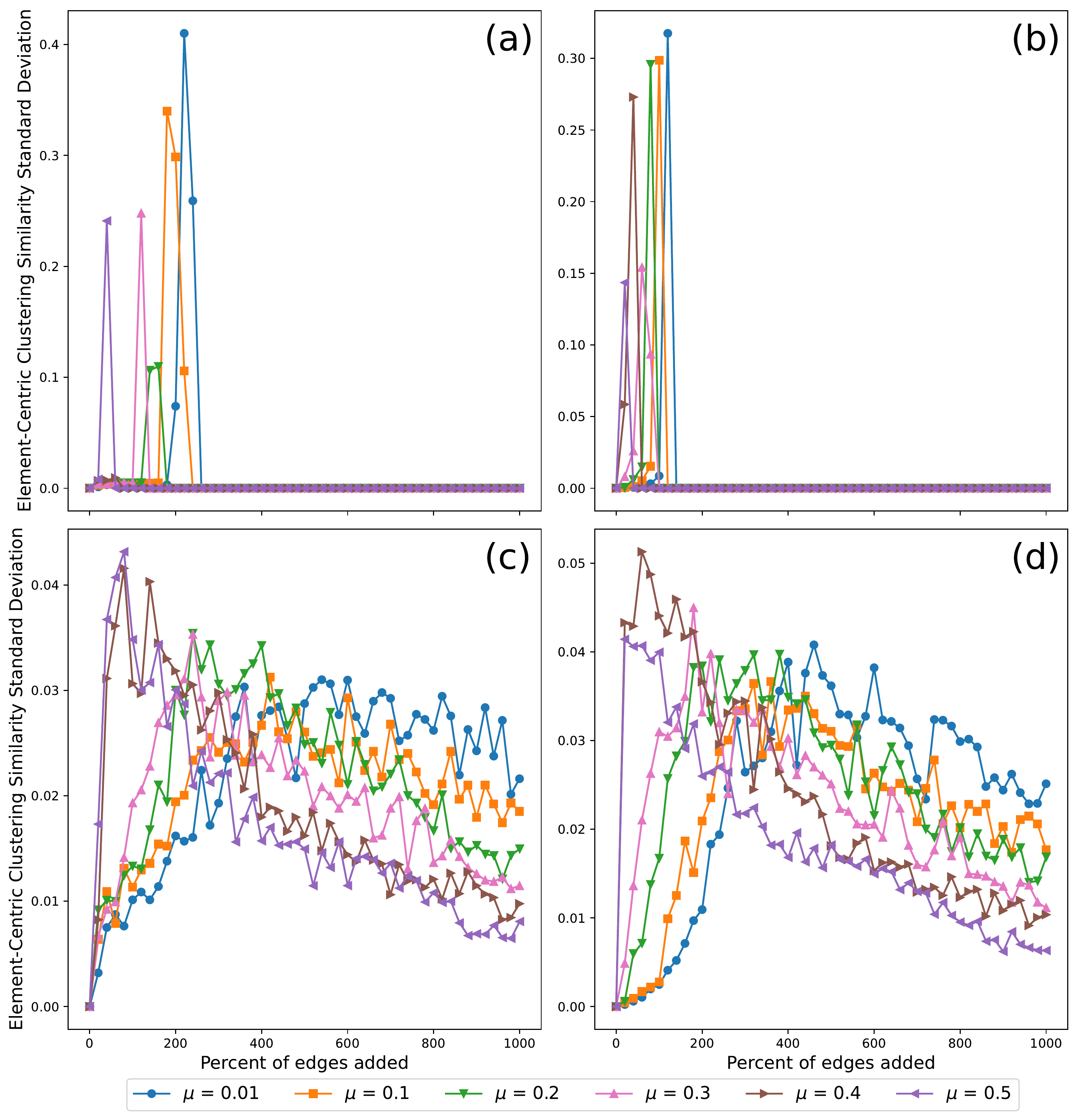}
\caption{\label{fig:LFR10000_elsimsd} Standard deviation of element-centric clustering similarity over the percentage of edges added uniformly at random on LFR benchmark graphs with $10\,000$ nodes. Communities detected by (a)~Infomap, (b)~Label Propagation, (c)~Leiden, and (d)~Louvain.}
\end{figure}

\begin{figure}[H]
\centering
\includegraphics[width=8.4cm]{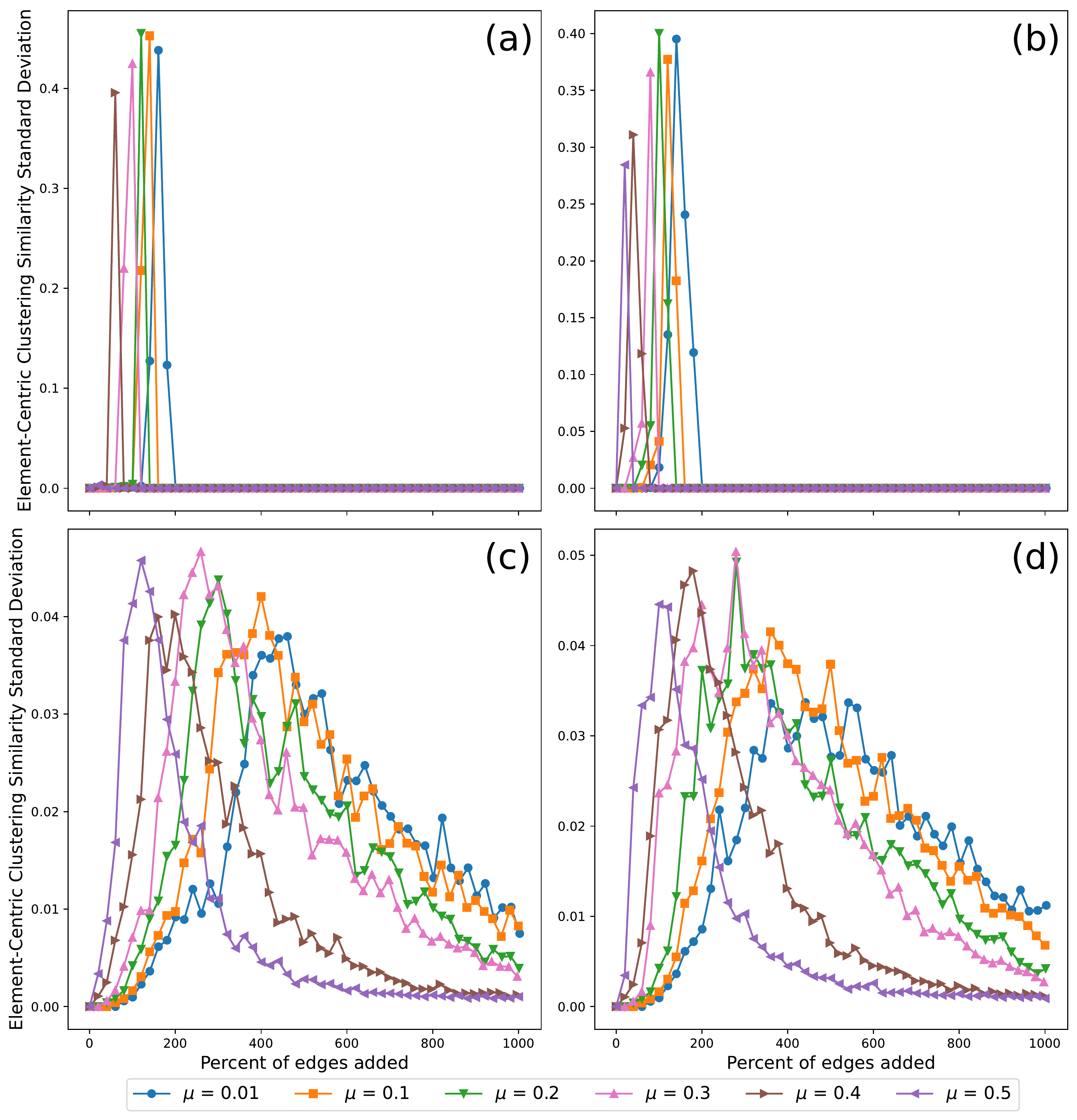}
\caption{\label{fig:LFR1000_elsimsd_target} Standard deviation of element-centric clustering similarity over the percentage of edges added that are selected uniformly at random across different communities on LFR benchmark graphs with $1000$ nodes. Communities detected by (a)~Infomap, (b)~Label Propagation, (c)~Leiden, and (d)~Louvain.}
\end{figure}

\begin{figure}[H]
\centering
\includegraphics[width=8.4cm]{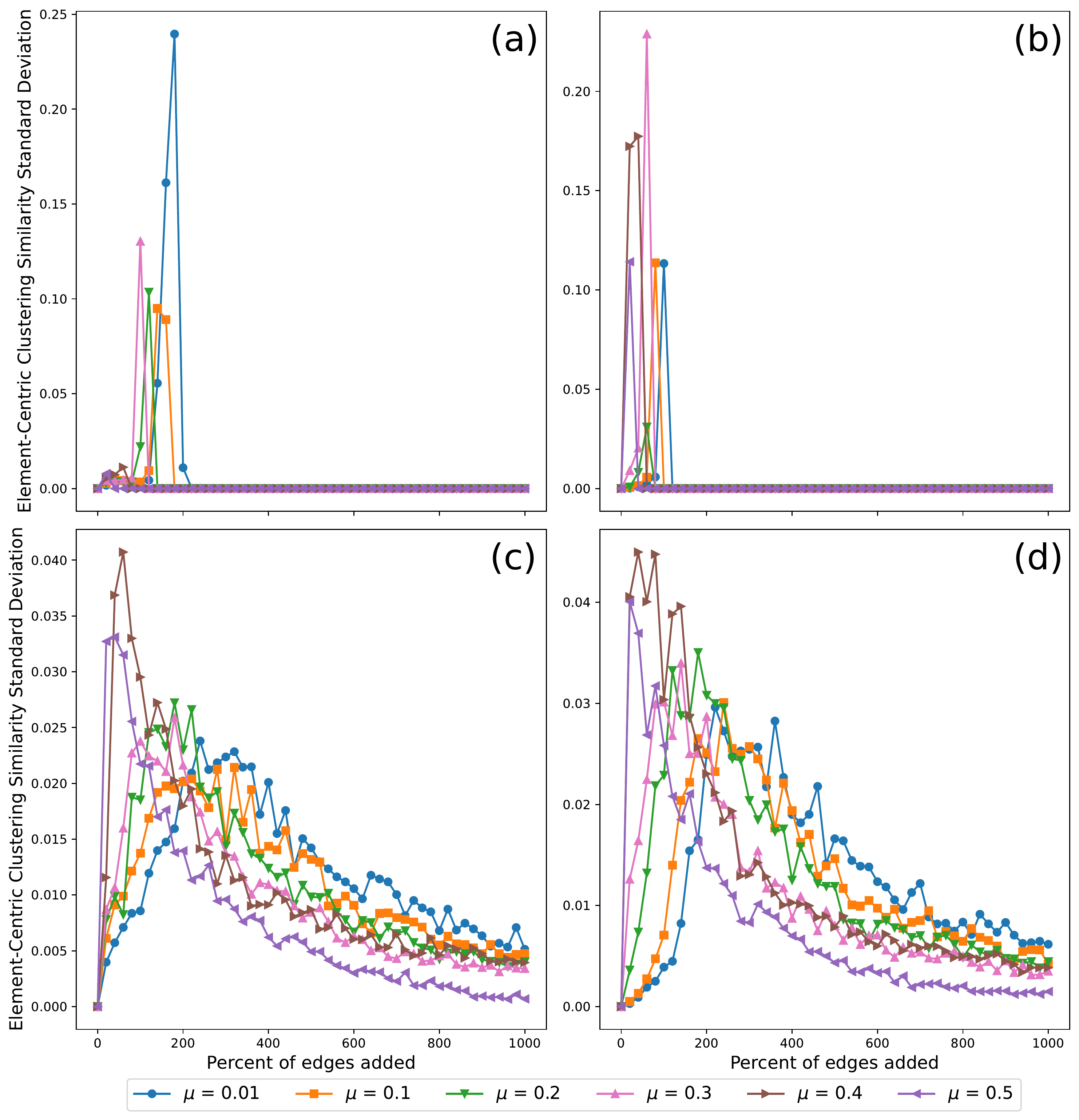}
\caption{\label{fig:LFR10000_elsimsd_target} Standard deviation of element-centric clustering similarity over the percentage of edges added that are selected uniformly at random across different communities on LFR benchmark graphs with $10\,000$ nodes. Communities detected by (a)~Infomap, (b)~Label Propagation, (c)~Leiden, and (d)~Louvain.}
\end{figure}

Figures~\ref{fig:LFR1000_nmisd} --~\ref{fig:LFR10000_elsimsd_target} show the standard deviation of NMI and element-centric clustering similarity metric for the synthetic results. Note that the overall changes in the standard deviation are generally restricted to a region with negligible scale compared with the mean. However, the only exception is for Infomap (a) and Label Propagation (b) where we observe a spike for almost every curve within the region when less than $2\times$ the original number of edges are added. Further investigation of the mean and the distributions reveals that these locations with large standard deviations correspond to the steps where the rapid drops in the means happen. Specifically, values of the community similarity metric at these steps are split into two families with comparable size -- one with $0$s and the other with values very close to $1$. Recall that at each step, there are $50$ independent realizations, meaning $50$ different perturbed graphs, so we suspect this is the uncertainty from the edge-addition stochastic process.

\newpage

\section{\label{app:emp-std} Standard Deviation for Empirical Results}

\begin{figure}[!h]
\centering
\includegraphics[width=7cm]{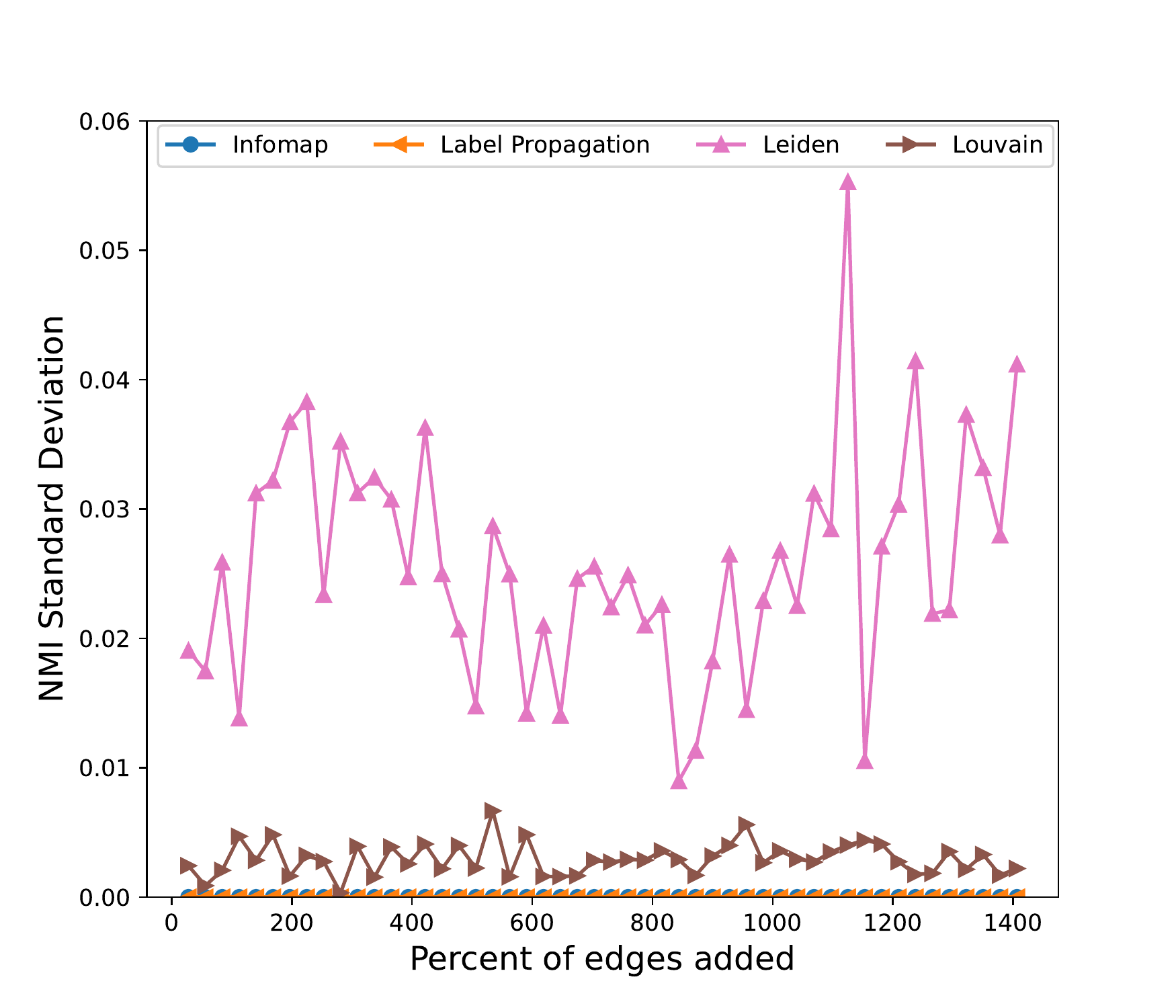}
\caption{\label{fig:rado_nmisd} Standard deviation of NMI over the percentage of edges added for the ia-radoslaw-email subnetwork.}
\end{figure}

\begin{figure}[h]
\centering
\includegraphics[width=7cm]{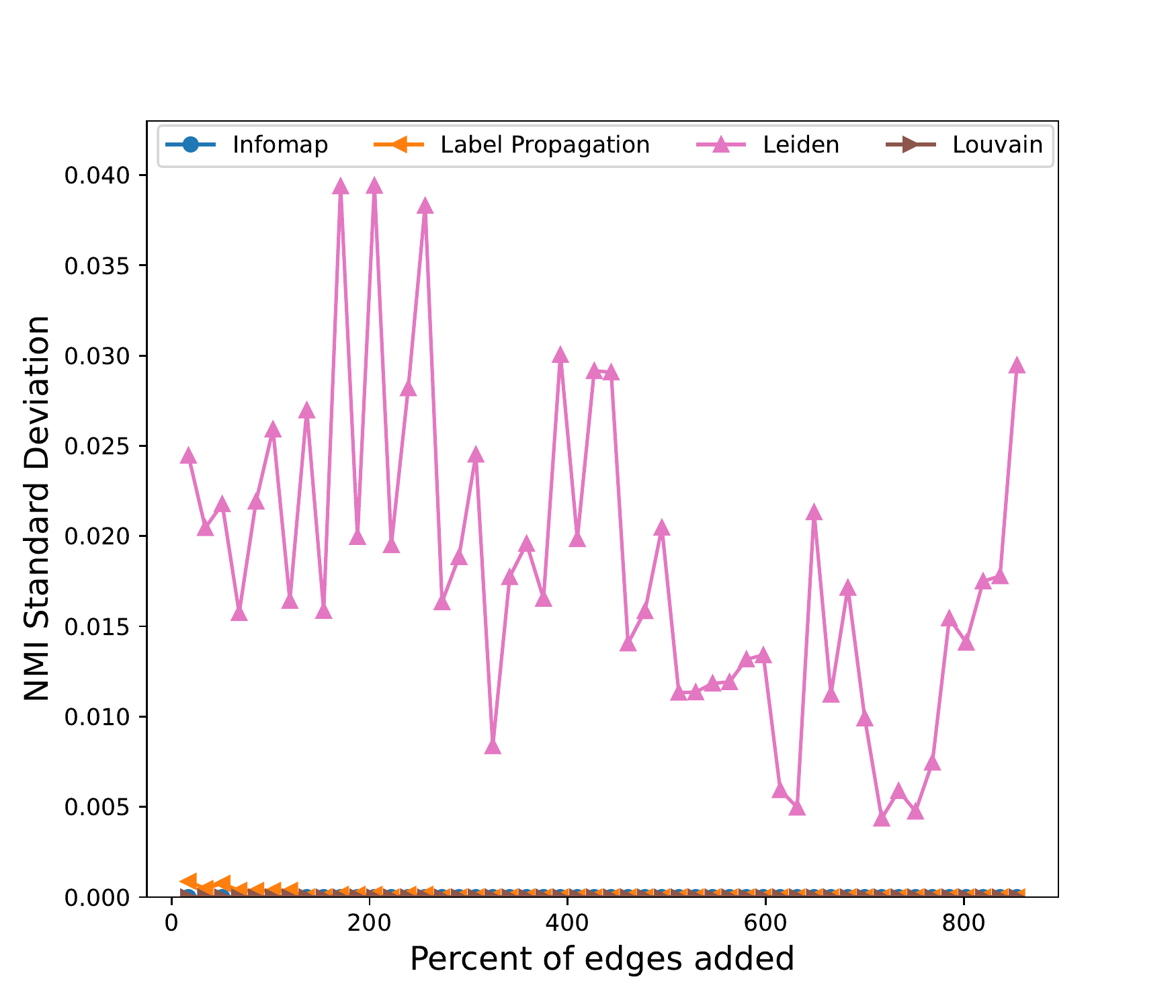}
\caption{\label{fig:enron_nmisd} Standard deviation of NMI over the percentage of edges added for the Enron subnetwork.}
\end{figure}

\begin{figure}[h]
\centering
\includegraphics[width=7cm]{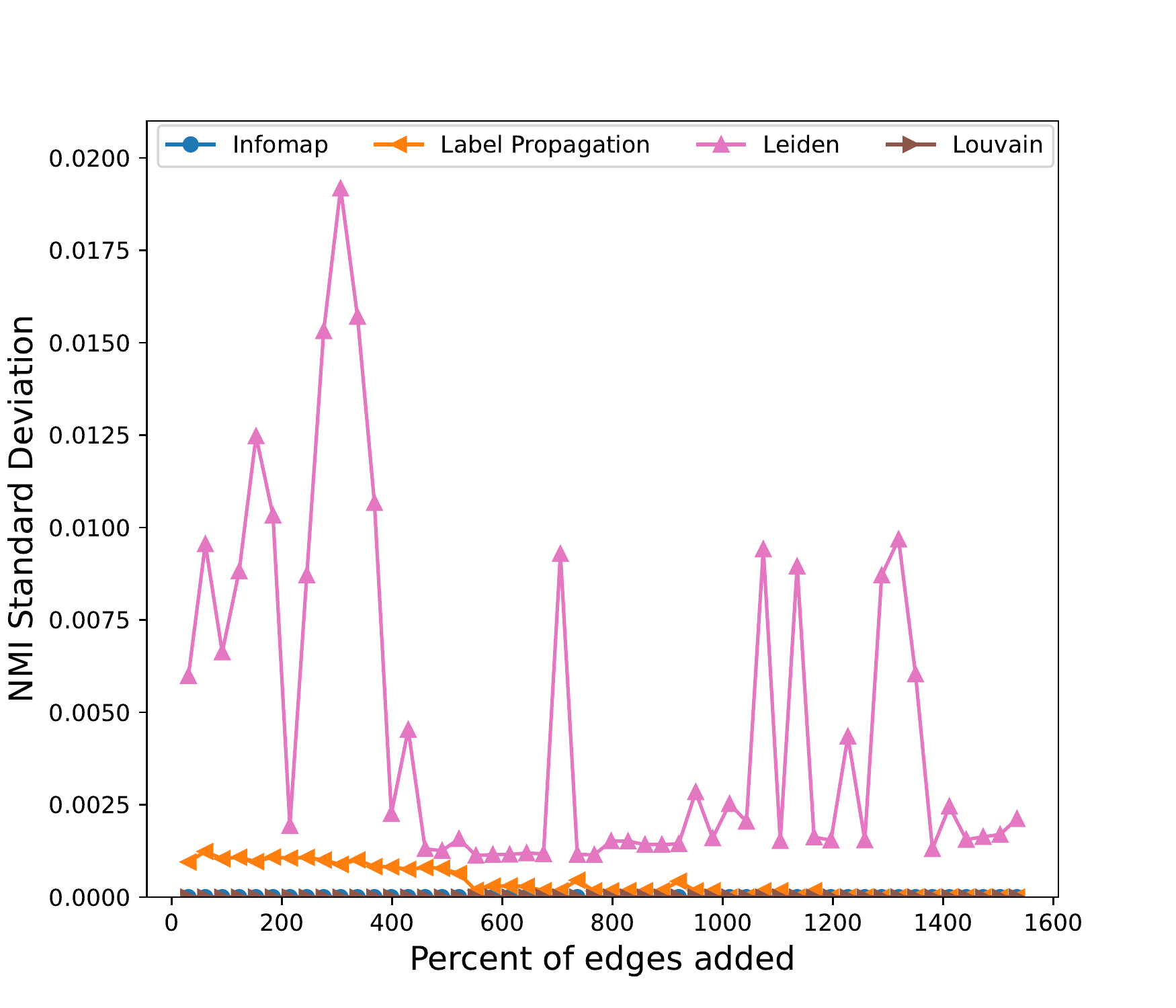}
\caption{\label{fig:eu_nmisd} Standard deviation of NMI over the percentage of edges added for the email-Eu-core-temporal subnetwork.}
\end{figure}

\begin{figure}[H]
\centering
\includegraphics[width=7cm]{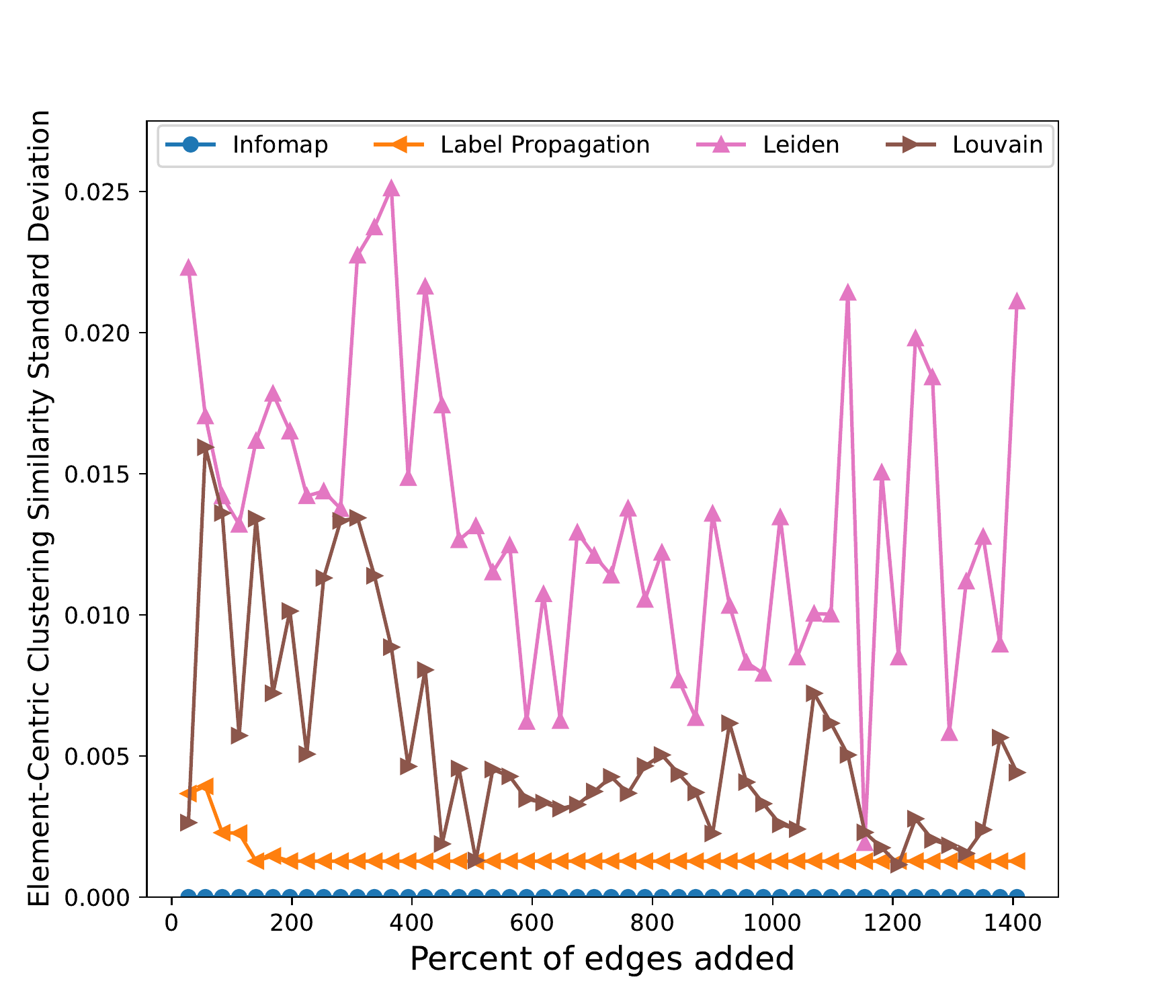}
\caption{\label{fig:rado_elsimsd} Standard deviation of element-centric clustering similarity over the percentage of edges added for the ia-radoslaw-email subnetwork.}
\end{figure}

\begin{figure}[h]
\centering
\includegraphics[width=7cm]{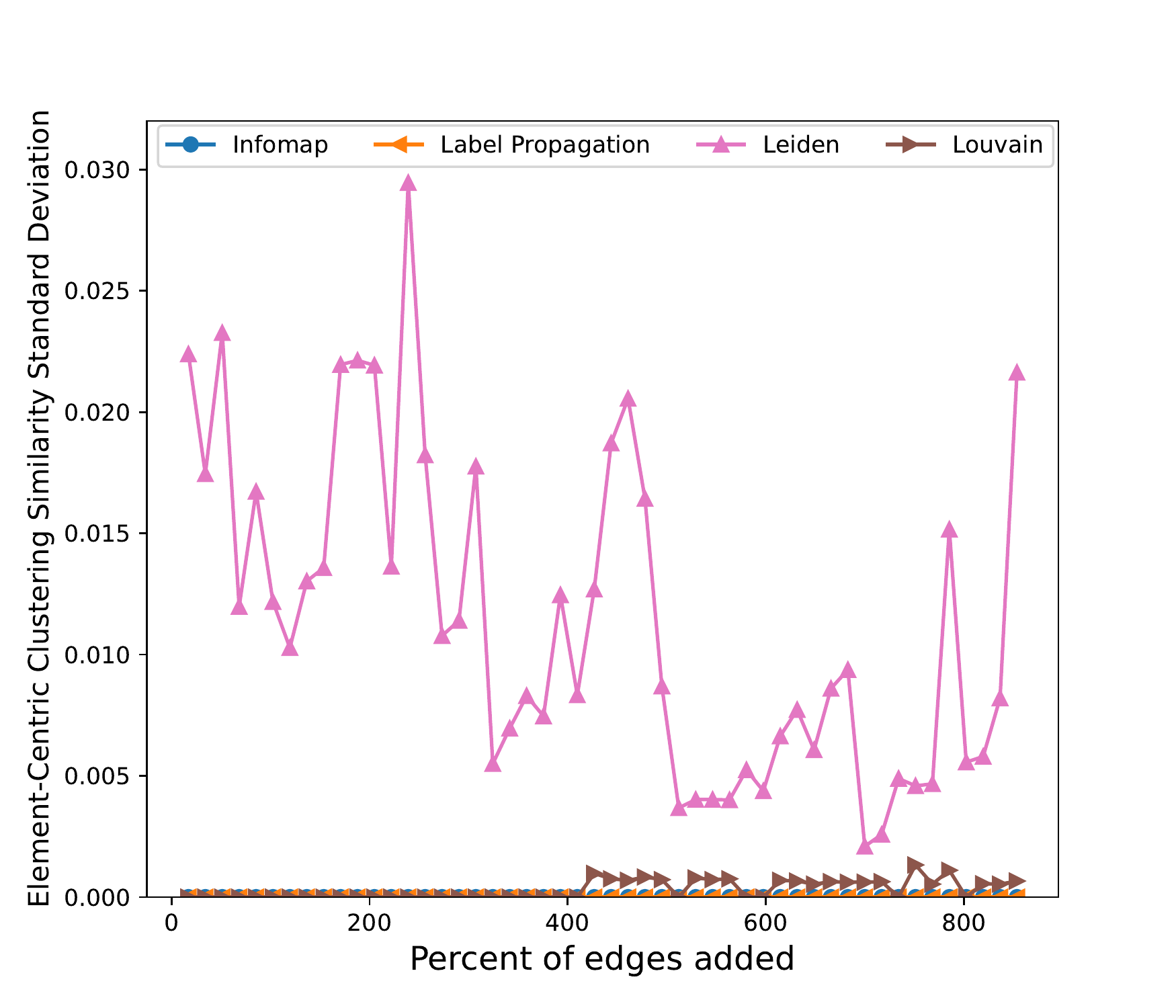}
\caption{\label{fig:enron_elsimsd} Standard deviation of element-centric clustering similarity over the percentage of edges added for the Enron subnetwork.}
\end{figure}

\begin{figure}[!t]
\centering
\includegraphics[width=7cm]{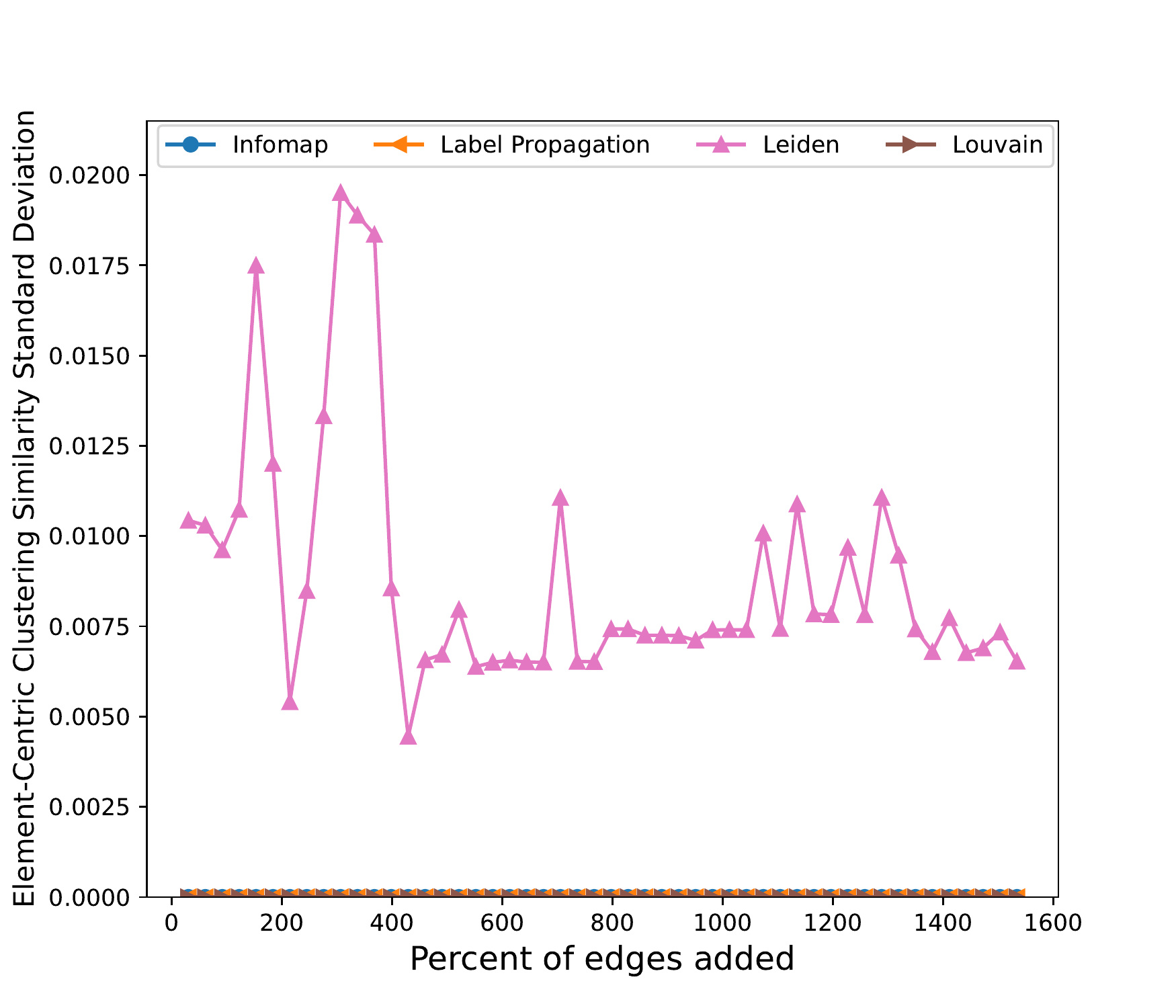}
\caption{\label{fig:eu_elsimsd} Standard deviation of element-centric clustering similarity over the percentage of edges added for the email-Eu-core-temporal subnetwork.}
\end{figure}

Figures~\ref{fig:rado_nmisd} --~\ref{fig:eu_elsimsd} show the standard deviation of NMI and element-centric clustering similarity metric for the empirical results. Note that the changes in standard deviation over steps are always restricted to a small region and the values are kept in scales that are negligible compared to the mean.

\clearpage
\bibliography{references}

\begin{thebibliography}{46}%
\makeatletter
\providecommand \@ifxundefined [1]{%
 \@ifx{#1\undefined}
}%
\providecommand \@ifnum [1]{%
 \ifnum #1\expandafter \@firstoftwo
 \else \expandafter \@secondoftwo
 \fi
}%
\providecommand \@ifx [1]{%
 \ifx #1\expandafter \@firstoftwo
 \else \expandafter \@secondoftwo
 \fi
}%
\providecommand \natexlab [1]{#1}%
\providecommand \enquote  [1]{``#1''}%
\providecommand \bibnamefont  [1]{#1}%
\providecommand \bibfnamefont [1]{#1}%
\providecommand \citenamefont [1]{#1}%
\providecommand \href@noop [0]{\@secondoftwo}%
\providecommand \href [0]{\begingroup \@sanitize@url \@href}%
\providecommand \@href[1]{\@@startlink{#1}\@@href}%
\providecommand \@@href[1]{\endgroup#1\@@endlink}%
\providecommand \@sanitize@url [0]{\catcode `\\12\catcode `\$12\catcode `\&12\catcode `\#12\catcode `\^12\catcode `\_12\catcode `\%12\relax}%
\providecommand \@@startlink[1]{}%
\providecommand \@@endlink[0]{}%
\providecommand \url  [0]{\begingroup\@sanitize@url \@url }%
\providecommand \@url [1]{\endgroup\@href {#1}{\urlprefix }}%
\providecommand \urlprefix  [0]{URL }%
\providecommand \Eprint [0]{\href }%
\providecommand \doibase [0]{https://doi.org/}%
\providecommand \selectlanguage [0]{\@gobble}%
\providecommand \bibinfo  [0]{\@secondoftwo}%
\providecommand \bibfield  [0]{\@secondoftwo}%
\providecommand \translation [1]{[#1]}%
\providecommand \BibitemOpen [0]{}%
\providecommand \bibitemStop [0]{}%
\providecommand \bibitemNoStop [0]{.\EOS\space}%
\providecommand \EOS [0]{\spacefactor3000\relax}%
\providecommand \BibitemShut  [1]{\csname bibitem#1\endcsname}%
\let\auto@bib@innerbib\@empty
\bibitem [{\citenamefont {Amaral}\ and\ \citenamefont {Ottino}(2004)}]{ComplexNetworks}%
  \BibitemOpen
  \bibfield  {author} {\bibinfo {author} {\bibfnamefont {L.~A.~N.}\ \bibnamefont {Amaral}}\ and\ \bibinfo {author} {\bibfnamefont {J.~M.}\ \bibnamefont {Ottino}},\ }\href {https://doi.org/10.1140/epjb/e2004-00110-5} {\bibfield  {journal} {\bibinfo  {journal} {Eur. Phys. J. B}\ }\textbf {\bibinfo {volume} {38}},\ \bibinfo {pages} {147} (\bibinfo {year} {2004})}\BibitemShut {NoStop}%
\bibitem [{\citenamefont {Newman}(2003)}]{NetworkStructure}%
  \BibitemOpen
  \bibfield  {author} {\bibinfo {author} {\bibfnamefont {M.~E.~J.}\ \bibnamefont {Newman}},\ }\href {https://doi.org/10.1137/S003614450342480} {\bibfield  {journal} {\bibinfo  {journal} {SIAM Rev.}\ }\textbf {\bibinfo {volume} {45}},\ \bibinfo {pages} {167} (\bibinfo {year} {2003})},\ \Eprint {https://arxiv.org/abs/https://doi.org/10.1137/S003614450342480} {https://doi.org/10.1137/S003614450342480} \BibitemShut {NoStop}%
\bibitem [{\citenamefont {Girvan}\ and\ \citenamefont {Newman}(2002)}]{GNBench}%
  \BibitemOpen
  \bibfield  {author} {\bibinfo {author} {\bibfnamefont {M.}~\bibnamefont {Girvan}}\ and\ \bibinfo {author} {\bibfnamefont {M.~E.~J.}\ \bibnamefont {Newman}},\ }\href {https://doi.org/10.1073/pnas.122653799} {\bibfield  {journal} {\bibinfo  {journal} {Proc. Natl. Acad. Sci. U.S.A.}\ }\textbf {\bibinfo {volume} {99}},\ \bibinfo {pages} {7821} (\bibinfo {year} {2002})},\ \Eprint {https://arxiv.org/abs/https://www.pnas.org/doi/pdf/10.1073/pnas.122653799} {https://www.pnas.org/doi/pdf/10.1073/pnas.122653799} \BibitemShut {NoStop}%
\bibitem [{\citenamefont {Fortunato}(2010)}]{Detection}%
  \BibitemOpen
  \bibfield  {author} {\bibinfo {author} {\bibfnamefont {S.}~\bibnamefont {Fortunato}},\ }\href {https://doi.org/https://doi.org/10.1016/j.physrep.2009.11.002} {\bibfield  {journal} {\bibinfo  {journal} {Phys. Rep.}\ }\textbf {\bibinfo {volume} {486}},\ \bibinfo {pages} {75} (\bibinfo {year} {2010})}\BibitemShut {NoStop}%
\bibitem [{\citenamefont {Fortunato}\ and\ \citenamefont {Hric}(2016)}]{Detection_UserGuide}%
  \BibitemOpen
  \bibfield  {author} {\bibinfo {author} {\bibfnamefont {S.}~\bibnamefont {Fortunato}}\ and\ \bibinfo {author} {\bibfnamefont {D.}~\bibnamefont {Hric}},\ }\href {https://doi.org/10.1016/j.physrep.2016.09.002} {\bibfield  {journal} {\bibinfo  {journal} {Phys. Rep.}\ }\textbf {\bibinfo {volume} {659}},\ \bibinfo {pages} {1} (\bibinfo {year} {2016})}\BibitemShut {NoStop}%
\bibitem [{\citenamefont {Karrer}\ \emph {et~al.}(2008)\citenamefont {Karrer}, \citenamefont {Levina},\ and\ \citenamefont {Newman}}]{robustKLN}%
  \BibitemOpen
  \bibfield  {author} {\bibinfo {author} {\bibfnamefont {B.}~\bibnamefont {Karrer}}, \bibinfo {author} {\bibfnamefont {E.}~\bibnamefont {Levina}},\ and\ \bibinfo {author} {\bibfnamefont {M.~E.~J.}\ \bibnamefont {Newman}},\ }\href {https://doi.org/10.1103/PhysRevE.77.046119} {\bibfield  {journal} {\bibinfo  {journal} {Phys. Rev. E}\ }\textbf {\bibinfo {volume} {77}},\ \bibinfo {pages} {046119} (\bibinfo {year} {2008})}\BibitemShut {NoStop}%
\bibitem [{\citenamefont {Barab{\'a}si}\ and\ \citenamefont {P{\'o}sfai}(2016)}]{barabasi_network}%
  \BibitemOpen
  \bibfield  {author} {\bibinfo {author} {\bibfnamefont {A.-L.}\ \bibnamefont {Barab{\'a}si}}\ and\ \bibinfo {author} {\bibfnamefont {M.}~\bibnamefont {P{\'o}sfai}},\ }\href {http://barabasi.com/networksciencebook/} {\emph {\bibinfo {title} {Network science}}}\ (\bibinfo  {publisher} {Cambridge University Press},\ \bibinfo {address} {Cambridge},\ \bibinfo {year} {2016})\BibitemShut {NoStop}%
\bibitem [{\citenamefont {Schaeffer}\ \emph {et~al.}(2021)\citenamefont {Schaeffer}, \citenamefont {Vald{\'e}s}, \citenamefont {Figols}, \citenamefont {Bachmann}, \citenamefont {Morales}, \citenamefont {Bustos-Jim{\'e}nez},\ and\ \citenamefont {Estrada}}]{RobustnessReview}%
  \BibitemOpen
  \bibfield  {author} {\bibinfo {author} {\bibfnamefont {S.~E.}\ \bibnamefont {Schaeffer}}, \bibinfo {author} {\bibfnamefont {V.}~\bibnamefont {Vald{\'e}s}}, \bibinfo {author} {\bibfnamefont {J.}~\bibnamefont {Figols}}, \bibinfo {author} {\bibfnamefont {I.}~\bibnamefont {Bachmann}}, \bibinfo {author} {\bibfnamefont {F.}~\bibnamefont {Morales}}, \bibinfo {author} {\bibfnamefont {J.}~\bibnamefont {Bustos-Jim{\'e}nez}},\ and\ \bibinfo {author} {\bibfnamefont {E.}~\bibnamefont {Estrada}},\ }\href {https://doi.org/10.1093/comnet/cnab018} {\bibfield  {journal} {\bibinfo  {journal} {J. Complex Netw.}\ }\textbf {\bibinfo {volume} {9}},\ \bibinfo {pages} {cnab018} (\bibinfo {year} {2021})}\BibitemShut {NoStop}%
\bibitem [{\citenamefont {Wang}\ and\ \citenamefont {Liu}(2019)}]{edge_attack}%
  \BibitemOpen
  \bibfield  {author} {\bibinfo {author} {\bibfnamefont {S.}~\bibnamefont {Wang}}\ and\ \bibinfo {author} {\bibfnamefont {J.}~\bibnamefont {Liu}},\ }\href {https://doi.org/10.1109/JSYST.2018.2835642} {\bibfield  {journal} {\bibinfo  {journal} {IEEE Syst. J.}\ }\textbf {\bibinfo {volume} {13}},\ \bibinfo {pages} {582} (\bibinfo {year} {2019})}\BibitemShut {NoStop}%
\bibitem [{\citenamefont {Albert}\ \emph {et~al.}(2000)\citenamefont {Albert}, \citenamefont {Jeong},\ and\ \citenamefont {Barab{\'a}si}}]{node_removal}%
  \BibitemOpen
  \bibfield  {author} {\bibinfo {author} {\bibfnamefont {R.}~\bibnamefont {Albert}}, \bibinfo {author} {\bibfnamefont {H.}~\bibnamefont {Jeong}},\ and\ \bibinfo {author} {\bibfnamefont {A.-L.}\ \bibnamefont {Barab{\'a}si}},\ }\href {https://doi.org/10.1038/35019019} {\bibfield  {journal} {\bibinfo  {journal} {Nature}\ }\textbf {\bibinfo {volume} {406}},\ \bibinfo {pages} {378} (\bibinfo {year} {2000})}\BibitemShut {NoStop}%
\bibitem [{\citenamefont {Amancio}\ \emph {et~al.}(2015)\citenamefont {Amancio}, \citenamefont {Oliveira},\ and\ \citenamefont {da~F~Costa}}]{robust_node_rm}%
  \BibitemOpen
  \bibfield  {author} {\bibinfo {author} {\bibfnamefont {D.~R.}\ \bibnamefont {Amancio}}, \bibinfo {author} {\bibfnamefont {O.~N.}\ \bibnamefont {Oliveira}},\ and\ \bibinfo {author} {\bibfnamefont {L.}~\bibnamefont {da~F~Costa}},\ }\href {https://doi.org/10.1088/1742-5468/2015/03/P03003} {\bibfield  {journal} {\bibinfo  {journal} {J. Stat. Mech.: Theory Exp.}\ }\textbf {\bibinfo {volume} {2015}}\bibinfo  {number} { (3)},\ \bibinfo {pages} {P03003}}\BibitemShut {NoStop}%
\bibitem [{\citenamefont {Wang}\ \emph {et~al.}(2017)\citenamefont {Wang}, \citenamefont {Liu},\ and\ \citenamefont {Wang}}]{node_removal_edge_rewiring}%
  \BibitemOpen
\bibfield  {number} {  }\bibfield  {author} {\bibinfo {author} {\bibfnamefont {S.}~\bibnamefont {Wang}}, \bibinfo {author} {\bibfnamefont {J.}~\bibnamefont {Liu}},\ and\ \bibinfo {author} {\bibfnamefont {X.}~\bibnamefont {Wang}},\ }\href {https://doi.org/10.1088/1742-5468/aa6581} {\bibfield  {journal} {\bibinfo  {journal} {J. Stat. Mech.: Theory Exp.}\ }\textbf {\bibinfo {volume} {2017}}\bibinfo  {number} { (4)},\ \bibinfo {pages} {043405}}\BibitemShut {NoStop}%
\bibitem [{\citenamefont {Carissimo}\ \emph {et~al.}(2018)\citenamefont {Carissimo}, \citenamefont {Cutillo},\ and\ \citenamefont {De~Feis}}]{robust_validation}%
  \BibitemOpen
\bibfield  {number} {  }\bibfield  {author} {\bibinfo {author} {\bibfnamefont {A.}~\bibnamefont {Carissimo}}, \bibinfo {author} {\bibfnamefont {L.}~\bibnamefont {Cutillo}},\ and\ \bibinfo {author} {\bibfnamefont {I.}~\bibnamefont {De~Feis}},\ }\href {https://doi.org/10.1016/j.csda.2017.10.006} {\bibfield  {journal} {\bibinfo  {journal} {Comput. Stat. Data Anal.}\ }\textbf {\bibinfo {volume} {120}},\ \bibinfo {pages} {1} (\bibinfo {year} {2018})}\BibitemShut {NoStop}%
\bibitem [{\citenamefont {Mozafari}\ and\ \citenamefont {Khansari}(2019)}]{improve_robust}%
  \BibitemOpen
  \bibfield  {author} {\bibinfo {author} {\bibfnamefont {M.}~\bibnamefont {Mozafari}}\ and\ \bibinfo {author} {\bibfnamefont {M.}~\bibnamefont {Khansari}},\ }\href {https://doi.org/10.1093/comnet/cnz009} {\bibfield  {journal} {\bibinfo  {journal} {J. Complex Netw.}\ }\textbf {\bibinfo {volume} {7}},\ \bibinfo {pages} {838} (\bibinfo {year} {2019})}\BibitemShut {NoStop}%
\bibitem [{\citenamefont {Leskovec}\ \emph {et~al.}(2005)\citenamefont {Leskovec}, \citenamefont {Kleinberg},\ and\ \citenamefont {Faloutsos}}]{DensifyingGraphs}%
  \BibitemOpen
  \bibfield  {author} {\bibinfo {author} {\bibfnamefont {J.}~\bibnamefont {Leskovec}}, \bibinfo {author} {\bibfnamefont {J.~M.}\ \bibnamefont {Kleinberg}},\ and\ \bibinfo {author} {\bibfnamefont {C.}~\bibnamefont {Faloutsos}},\ }in\ \href@noop {} {\emph {\bibinfo {booktitle} {Proceedings of the Eleventh ACM SIGKDD International Conference on Knowledge Discovery in Data Mining}}},\ \bibinfo {organization} {KDD '05}\ (\bibinfo  {publisher} {ACM},\ \bibinfo {address} {New York},\ \bibinfo {year} {2005})\ pp.\ \bibinfo {pages} {177--187}\BibitemShut {NoStop}%
\bibitem [{\citenamefont {Holme}\ \emph {et~al.}(2002)\citenamefont {Holme}, \citenamefont {Kim}, \citenamefont {Yoon},\ and\ \citenamefont {Han}}]{attack}%
  \BibitemOpen
  \bibfield  {author} {\bibinfo {author} {\bibfnamefont {P.}~\bibnamefont {Holme}}, \bibinfo {author} {\bibfnamefont {B.~J.}\ \bibnamefont {Kim}}, \bibinfo {author} {\bibfnamefont {C.~N.}\ \bibnamefont {Yoon}},\ and\ \bibinfo {author} {\bibfnamefont {S.~K.}\ \bibnamefont {Han}},\ }\href {https://doi.org/10.1103/PhysRevE.65.056109} {\bibfield  {journal} {\bibinfo  {journal} {Phys. Rev. E}\ }\textbf {\bibinfo {volume} {65}},\ \bibinfo {pages} {056109} (\bibinfo {year} {2002})}\BibitemShut {NoStop}%
\bibitem [{\citenamefont {Borgatti}\ \emph {et~al.}(2006)\citenamefont {Borgatti}, \citenamefont {Carley},\ and\ \citenamefont {Krackhardt}}]{addition_error}%
  \BibitemOpen
  \bibfield  {author} {\bibinfo {author} {\bibfnamefont {S.~P.}\ \bibnamefont {Borgatti}}, \bibinfo {author} {\bibfnamefont {K.~M.}\ \bibnamefont {Carley}},\ and\ \bibinfo {author} {\bibfnamefont {D.}~\bibnamefont {Krackhardt}},\ }\href {https://doi.org/https://doi.org/10.1016/j.socnet.2005.05.001} {\bibfield  {journal} {\bibinfo  {journal} {Soc. Netw.}\ }\textbf {\bibinfo {volume} {28}},\ \bibinfo {pages} {124} (\bibinfo {year} {2006})}\BibitemShut {NoStop}%
\bibitem [{\citenamefont {Wang}\ \emph {et~al.}(2012)\citenamefont {Wang}, \citenamefont {Shi}, \citenamefont {McFarland},\ and\ \citenamefont {Leskovec}}]{measurement_error}%
  \BibitemOpen
  \bibfield  {author} {\bibinfo {author} {\bibfnamefont {D.~J.}\ \bibnamefont {Wang}}, \bibinfo {author} {\bibfnamefont {X.}~\bibnamefont {Shi}}, \bibinfo {author} {\bibfnamefont {D.~A.}\ \bibnamefont {McFarland}},\ and\ \bibinfo {author} {\bibfnamefont {J.}~\bibnamefont {Leskovec}},\ }\href@noop {} {\bibfield  {journal} {\bibinfo  {journal} {Soc. Netw.}\ }\textbf {\bibinfo {volume} {34}},\ \bibinfo {pages} {396} (\bibinfo {year} {2012})}\BibitemShut {NoStop}%
\bibitem [{\citenamefont {Zargar}\ \emph {et~al.}(2013)\citenamefont {Zargar}, \citenamefont {Joshi},\ and\ \citenamefont {Tipper}}]{DoS1}%
  \BibitemOpen
  \bibfield  {author} {\bibinfo {author} {\bibfnamefont {S.~T.}\ \bibnamefont {Zargar}}, \bibinfo {author} {\bibfnamefont {J.}~\bibnamefont {Joshi}},\ and\ \bibinfo {author} {\bibfnamefont {D.}~\bibnamefont {Tipper}},\ }\href {https://doi.org/10.1109/SURV.2013.031413.00127} {\bibfield  {journal} {\bibinfo  {journal} {IEEE Commun. Surv. Tutor.}\ }\textbf {\bibinfo {volume} {15}},\ \bibinfo {pages} {2046} (\bibinfo {year} {2013})}\BibitemShut {NoStop}%
\bibitem [{\citenamefont {Osanaiye}\ \emph {et~al.}(2016)\citenamefont {Osanaiye}, \citenamefont {Choo},\ and\ \citenamefont {Dlodlo}}]{DoS2}%
  \BibitemOpen
  \bibfield  {author} {\bibinfo {author} {\bibfnamefont {O.}~\bibnamefont {Osanaiye}}, \bibinfo {author} {\bibfnamefont {K.-K.~R.}\ \bibnamefont {Choo}},\ and\ \bibinfo {author} {\bibfnamefont {M.}~\bibnamefont {Dlodlo}},\ }\href {https://doi.org/https://doi.org/10.1016/j.jnca.2016.01.001} {\bibfield  {journal} {\bibinfo  {journal} {J. Netw. Comput. Appl.}\ }\textbf {\bibinfo {volume} {67}},\ \bibinfo {pages} {147} (\bibinfo {year} {2016})}\BibitemShut {NoStop}%
\bibitem [{\citenamefont {Lancichinetti}\ and\ \citenamefont {Fortunato}(2009{\natexlab{a}})}]{LFRBenchNew}%
  \BibitemOpen
  \bibfield  {author} {\bibinfo {author} {\bibfnamefont {A.}~\bibnamefont {Lancichinetti}}\ and\ \bibinfo {author} {\bibfnamefont {S.}~\bibnamefont {Fortunato}},\ }\href {https://doi.org/10.1103/PhysRevE.80.016118} {\bibfield  {journal} {\bibinfo  {journal} {Phys. Rev. E}\ }\textbf {\bibinfo {volume} {80}},\ \bibinfo {pages} {016118} (\bibinfo {year} {2009}{\natexlab{a}})}\BibitemShut {NoStop}%
\bibitem [{\citenamefont {Fred}\ and\ \citenamefont {Jain}(2003)}]{NMI}%
  \BibitemOpen
  \bibfield  {author} {\bibinfo {author} {\bibfnamefont {A.~L.~N.}\ \bibnamefont {Fred}}\ and\ \bibinfo {author} {\bibfnamefont {A.~K.}\ \bibnamefont {Jain}},\ }\href@noop {} {\bibfield  {journal} {\bibinfo  {journal} {2003 Proc. IEEE Comput. Soc. Conf. Comput. Vis. Pattern Recognit.}\ }\textbf {\bibinfo {volume} {2}},\ \bibinfo {pages} {II} (\bibinfo {year} {2003})}\BibitemShut {NoStop}%
\bibitem [{\citenamefont {Gates}\ \emph {et~al.}(2019)\citenamefont {Gates}, \citenamefont {Wood}, \citenamefont {Hetrick},\ and\ \citenamefont {Ahn}}]{element-centric}%
  \BibitemOpen
  \bibfield  {author} {\bibinfo {author} {\bibfnamefont {A.~J.}\ \bibnamefont {Gates}}, \bibinfo {author} {\bibfnamefont {I.~B.}\ \bibnamefont {Wood}}, \bibinfo {author} {\bibfnamefont {W.~P.}\ \bibnamefont {Hetrick}},\ and\ \bibinfo {author} {\bibfnamefont {Y.-Y.}\ \bibnamefont {Ahn}},\ }\href {https://doi.org/10.1038/s41598-019-44892-y} {\bibfield  {journal} {\bibinfo  {journal} {Sci. Rep.}\ }\textbf {\bibinfo {volume} {9}},\ \bibinfo {pages} {8574} (\bibinfo {year} {2019})}\BibitemShut {NoStop}%
\bibitem [{\citenamefont {Lancichinetti}\ and\ \citenamefont {Fortunato}(2009{\natexlab{b}})}]{ClusterAlg}%
  \BibitemOpen
  \bibfield  {author} {\bibinfo {author} {\bibfnamefont {A.}~\bibnamefont {Lancichinetti}}\ and\ \bibinfo {author} {\bibfnamefont {S.}~\bibnamefont {Fortunato}},\ }\href {https://doi.org/10.1103/PhysRevE.80.056117} {\bibfield  {journal} {\bibinfo  {journal} {Phys. Rev. E}\ }\textbf {\bibinfo {volume} {80}},\ \bibinfo {pages} {056117} (\bibinfo {year} {2009}{\natexlab{b}})}\BibitemShut {NoStop}%
\bibitem [{\citenamefont {Emmons}\ \emph {et~al.}(2016)\citenamefont {Emmons}, \citenamefont {Kobourov}, \citenamefont {Gallant},\ and\ \citenamefont {B{\"o}rner}}]{AnaCluQua}%
  \BibitemOpen
  \bibfield  {author} {\bibinfo {author} {\bibfnamefont {S.}~\bibnamefont {Emmons}}, \bibinfo {author} {\bibfnamefont {S.}~\bibnamefont {Kobourov}}, \bibinfo {author} {\bibfnamefont {M.}~\bibnamefont {Gallant}},\ and\ \bibinfo {author} {\bibfnamefont {K.}~\bibnamefont {B{\"o}rner}},\ }\href {https://doi.org/10.1371/journal.pone.0159161} {\bibfield  {journal} {\bibinfo  {journal} {PLOS ONE}\ }\textbf {\bibinfo {volume} {11}},\ \bibinfo {pages} {1} (\bibinfo {year} {2016})}\BibitemShut {NoStop}%
\bibitem [{\citenamefont {Lancichinetti}\ \emph {et~al.}(2008)\citenamefont {Lancichinetti}, \citenamefont {Fortunato},\ and\ \citenamefont {Radicchi}}]{LFRBench}%
  \BibitemOpen
  \bibfield  {author} {\bibinfo {author} {\bibfnamefont {A.}~\bibnamefont {Lancichinetti}}, \bibinfo {author} {\bibfnamefont {S.}~\bibnamefont {Fortunato}},\ and\ \bibinfo {author} {\bibfnamefont {F.}~\bibnamefont {Radicchi}},\ }\href {https://doi.org/10.1103/PhysRevE.78.046110} {\bibfield  {journal} {\bibinfo  {journal} {Phys. Rev. E}\ }\textbf {\bibinfo {volume} {78}},\ \bibinfo {pages} {046110} (\bibinfo {year} {2008})}\BibitemShut {NoStop}%
\bibitem [{\citenamefont {Palla}\ \emph {et~al.}(2005)\citenamefont {Palla}, \citenamefont {Der{\'e}nyi}, \citenamefont {Farkas},\ and\ \citenamefont {Vicsek}}]{power_law}%
  \BibitemOpen
  \bibfield  {author} {\bibinfo {author} {\bibfnamefont {G.}~\bibnamefont {Palla}}, \bibinfo {author} {\bibfnamefont {I.}~\bibnamefont {Der{\'e}nyi}}, \bibinfo {author} {\bibfnamefont {I.}~\bibnamefont {Farkas}},\ and\ \bibinfo {author} {\bibfnamefont {T.}~\bibnamefont {Vicsek}},\ }\href {https://doi.org/10.1038/nature03607} {\bibfield  {journal} {\bibinfo  {journal} {Nature}\ }\textbf {\bibinfo {volume} {435}},\ \bibinfo {pages} {814} (\bibinfo {year} {2005})}\BibitemShut {NoStop}%
\bibitem [{\citenamefont {Clauset}\ \emph {et~al.}(2004)\citenamefont {Clauset}, \citenamefont {Newman},\ and\ \citenamefont {Moore}}]{power_law2}%
  \BibitemOpen
  \bibfield  {author} {\bibinfo {author} {\bibfnamefont {A.}~\bibnamefont {Clauset}}, \bibinfo {author} {\bibfnamefont {M.~E.~J.}\ \bibnamefont {Newman}},\ and\ \bibinfo {author} {\bibfnamefont {C.}~\bibnamefont {Moore}},\ }\href {https://doi.org/10.1103/PhysRevE.70.066111} {\bibfield  {journal} {\bibinfo  {journal} {Phys. Rev. E}\ }\textbf {\bibinfo {volume} {70}},\ \bibinfo {pages} {066111} (\bibinfo {year} {2004})}\BibitemShut {NoStop}%
\bibitem [{\citenamefont {Guimer\`a}\ \emph {et~al.}(2003)\citenamefont {Guimer\`a}, \citenamefont {Danon}, \citenamefont {D\'{\i}az-Guilera}, \citenamefont {Giralt},\ and\ \citenamefont {Arenas}}]{power_law3}%
  \BibitemOpen
  \bibfield  {author} {\bibinfo {author} {\bibfnamefont {R.}~\bibnamefont {Guimer\`a}}, \bibinfo {author} {\bibfnamefont {L.}~\bibnamefont {Danon}}, \bibinfo {author} {\bibfnamefont {A.}~\bibnamefont {D\'{\i}az-Guilera}}, \bibinfo {author} {\bibfnamefont {F.}~\bibnamefont {Giralt}},\ and\ \bibinfo {author} {\bibfnamefont {A.}~\bibnamefont {Arenas}},\ }\href {https://doi.org/10.1103/PhysRevE.68.065103} {\bibfield  {journal} {\bibinfo  {journal} {Phys. Rev. E}\ }\textbf {\bibinfo {volume} {68}},\ \bibinfo {pages} {065103(R)} (\bibinfo {year} {2003})}\BibitemShut {NoStop}%
\bibitem [{\citenamefont {Holland}\ \emph {et~al.}(1983)\citenamefont {Holland}, \citenamefont {Laskey},\ and\ \citenamefont {Leinhardt}}]{SBM_original}%
  \BibitemOpen
  \bibfield  {author} {\bibinfo {author} {\bibfnamefont {P.~W.}\ \bibnamefont {Holland}}, \bibinfo {author} {\bibfnamefont {K.~B.}\ \bibnamefont {Laskey}},\ and\ \bibinfo {author} {\bibfnamefont {S.}~\bibnamefont {Leinhardt}},\ }\href {https://doi.org/https://doi.org/10.1016/0378-8733(83)90021-7} {\bibfield  {journal} {\bibinfo  {journal} {Soc. Netw.}\ }\textbf {\bibinfo {volume} {5}},\ \bibinfo {pages} {109} (\bibinfo {year} {1983})}\BibitemShut {NoStop}%
\bibitem [{\citenamefont {Fortunato}(2023)}]{Fortunato_website}%
  \BibitemOpen
  \bibfield  {author} {\bibinfo {author} {\bibfnamefont {S.}~\bibnamefont {Fortunato}},\ }\href@noop {} {\bibinfo {title} {Santo \uppercase{F}ortunato's \uppercase{W}ebsite: resources}},\ \bibinfo {howpublished} {https://www.santofortunato.net/resources} (\bibinfo {year} {Last Accessed: January 11, 2023})\BibitemShut {NoStop}%
\bibitem [{\citenamefont {Rosvall}\ and\ \citenamefont {Bergstrom}(2008)}]{infomap}%
  \BibitemOpen
  \bibfield  {author} {\bibinfo {author} {\bibfnamefont {M.}~\bibnamefont {Rosvall}}\ and\ \bibinfo {author} {\bibfnamefont {C.~T.}\ \bibnamefont {Bergstrom}},\ }\href {https://doi.org/10.1073/pnas.0706851105} {\bibfield  {journal} {\bibinfo  {journal} {Proc. Natl. Acad. Sci. U.S.A.}\ }\textbf {\bibinfo {volume} {105}},\ \bibinfo {pages} {1118} (\bibinfo {year} {2008})},\ \Eprint {https://arxiv.org/abs/https://www.pnas.org/doi/pdf/10.1073/pnas.0706851105} {https://www.pnas.org/doi/pdf/10.1073/pnas.0706851105} \BibitemShut {NoStop}%
\bibitem [{\citenamefont {Raghavan}\ \emph {et~al.}(2007)\citenamefont {Raghavan}, \citenamefont {Albert},\ and\ \citenamefont {Kumara}}]{lpm}%
  \BibitemOpen
  \bibfield  {author} {\bibinfo {author} {\bibfnamefont {U.~N.}\ \bibnamefont {Raghavan}}, \bibinfo {author} {\bibfnamefont {R.}~\bibnamefont {Albert}},\ and\ \bibinfo {author} {\bibfnamefont {S.}~\bibnamefont {Kumara}},\ }\href {https://doi.org/10.1103/PhysRevE.76.036106} {\bibfield  {journal} {\bibinfo  {journal} {Phys. Rev. E}\ }\textbf {\bibinfo {volume} {76}},\ \bibinfo {pages} {036106} (\bibinfo {year} {2007})}\BibitemShut {NoStop}%
\bibitem [{\citenamefont {Traag}\ \emph {et~al.}(2019)\citenamefont {Traag}, \citenamefont {Waltman},\ and\ \citenamefont {van Eck}}]{leiden}%
  \BibitemOpen
  \bibfield  {author} {\bibinfo {author} {\bibfnamefont {V.~A.}\ \bibnamefont {Traag}}, \bibinfo {author} {\bibfnamefont {L.}~\bibnamefont {Waltman}},\ and\ \bibinfo {author} {\bibfnamefont {N.~J.}\ \bibnamefont {van Eck}},\ }\href {https://doi.org/10.1038/s41598-019-41695-z} {\bibfield  {journal} {\bibinfo  {journal} {Sci. Rep.}\ }\textbf {\bibinfo {volume} {9}},\ \bibinfo {pages} {5233} (\bibinfo {year} {2019})}\BibitemShut {NoStop}%
\bibitem [{\citenamefont {Blondel}\ \emph {et~al.}(2008)\citenamefont {Blondel}, \citenamefont {Guillaume}, \citenamefont {Lambiotte},\ and\ \citenamefont {Lefebvre}}]{louvain}%
  \BibitemOpen
  \bibfield  {author} {\bibinfo {author} {\bibfnamefont {V.~D.}\ \bibnamefont {Blondel}}, \bibinfo {author} {\bibfnamefont {J.-L.}\ \bibnamefont {Guillaume}}, \bibinfo {author} {\bibfnamefont {R.}~\bibnamefont {Lambiotte}},\ and\ \bibinfo {author} {\bibfnamefont {E.}~\bibnamefont {Lefebvre}},\ }\href {https://doi.org/10.1088/1742-5468/2008/10/p10008} {\bibfield  {journal} {\bibinfo  {journal} {J. Stat. Mech.: Theory Exp.}\ }\textbf {\bibinfo {volume} {2008}}\bibinfo  {number} { (10)},\ \bibinfo {pages} {P10008}}\BibitemShut {NoStop}%
\bibitem [{\citenamefont {Yang}\ \emph {et~al.}(2016)\citenamefont {Yang}, \citenamefont {Algesheimer},\ and\ \citenamefont {Tessone}}]{detection_analysis}%
  \BibitemOpen
\bibfield  {number} {  }\bibfield  {author} {\bibinfo {author} {\bibfnamefont {Z.}~\bibnamefont {Yang}}, \bibinfo {author} {\bibfnamefont {R.}~\bibnamefont {Algesheimer}},\ and\ \bibinfo {author} {\bibfnamefont {C.~J.}\ \bibnamefont {Tessone}},\ }\href {https://doi.org/10.1038/srep30750} {\bibfield  {journal} {\bibinfo  {journal} {Sci. Rep.}\ }\textbf {\bibinfo {volume} {6}},\ \bibinfo {pages} {30750} (\bibinfo {year} {2016})}\BibitemShut {NoStop}%
\bibitem [{\citenamefont {Lancichinetti}\ and\ \citenamefont {Fortunato}(2012)}]{ConsensusClu}%
  \BibitemOpen
  \bibfield  {author} {\bibinfo {author} {\bibfnamefont {A.}~\bibnamefont {Lancichinetti}}\ and\ \bibinfo {author} {\bibfnamefont {S.}~\bibnamefont {Fortunato}},\ }\href {https://doi.org/10.1038/srep00336} {\bibfield  {journal} {\bibinfo  {journal} {Sci. Rep.}\ }\textbf {\bibinfo {volume} {2}},\ \bibinfo {pages} {336} (\bibinfo {year} {2012})}\BibitemShut {NoStop}%
\bibitem [{\citenamefont {Traag}(2023)}]{leiden_website}%
  \BibitemOpen
  \bibfield  {author} {\bibinfo {author} {\bibfnamefont {V.~A.}\ \bibnamefont {Traag}},\ }\href@noop {} {\bibinfo {title} {Github repository: networkanalysis}},\ \bibinfo {howpublished} {https://github.com/CWTSLeiden/networkanalysis} (\bibinfo {year} {Last Accessed: January 11, 2023})\BibitemShut {NoStop}%
\bibitem [{\citenamefont {Tandon}\ \emph {et~al.}(2021)\citenamefont {Tandon}, \citenamefont {Albeshri}, \citenamefont {Thayananthan}, \citenamefont {Alhalabi}, \citenamefont {Radicchi},\ and\ \citenamefont {Fortunato}}]{CommDetect_Embedding}%
  \BibitemOpen
  \bibfield  {author} {\bibinfo {author} {\bibfnamefont {A.}~\bibnamefont {Tandon}}, \bibinfo {author} {\bibfnamefont {A.}~\bibnamefont {Albeshri}}, \bibinfo {author} {\bibfnamefont {V.}~\bibnamefont {Thayananthan}}, \bibinfo {author} {\bibfnamefont {W.}~\bibnamefont {Alhalabi}}, \bibinfo {author} {\bibfnamefont {F.}~\bibnamefont {Radicchi}},\ and\ \bibinfo {author} {\bibfnamefont {S.}~\bibnamefont {Fortunato}},\ }\href {https://doi.org/10.1103/PhysRevE.103.022316} {\bibfield  {journal} {\bibinfo  {journal} {Phys. Rev. E}\ }\textbf {\bibinfo {volume} {103}},\ \bibinfo {pages} {022316} (\bibinfo {year} {2021})}\BibitemShut {NoStop}%
\bibitem [{\citenamefont {Pedregosa}\ \emph {et~al.}(2011)\citenamefont {Pedregosa}, \citenamefont {Varoquaux}, \citenamefont {Gramfort}, \citenamefont {Michel}, \citenamefont {Thirion}, \citenamefont {Grisel}, \citenamefont {Blondel}, \citenamefont {Prettenhofer}, \citenamefont {Weiss}, \citenamefont {Dubourg}, \citenamefont {Vanderplas}, \citenamefont {Passos}, \citenamefont {Cournapeau}, \citenamefont {Brucher}, \citenamefont {Perrot},\ and\ \citenamefont {Duchesnay}}]{scikit-learn}%
  \BibitemOpen
  \bibfield  {author} {\bibinfo {author} {\bibfnamefont {F.}~\bibnamefont {Pedregosa}}, \bibinfo {author} {\bibfnamefont {G.}~\bibnamefont {Varoquaux}}, \bibinfo {author} {\bibfnamefont {A.}~\bibnamefont {Gramfort}}, \bibinfo {author} {\bibfnamefont {V.}~\bibnamefont {Michel}}, \bibinfo {author} {\bibfnamefont {B.}~\bibnamefont {Thirion}}, \bibinfo {author} {\bibfnamefont {O.}~\bibnamefont {Grisel}}, \bibinfo {author} {\bibfnamefont {M.}~\bibnamefont {Blondel}}, \bibinfo {author} {\bibfnamefont {P.}~\bibnamefont {Prettenhofer}}, \bibinfo {author} {\bibfnamefont {R.}~\bibnamefont {Weiss}}, \bibinfo {author} {\bibfnamefont {V.}~\bibnamefont {Dubourg}}, \bibinfo {author} {\bibfnamefont {J.}~\bibnamefont {Vanderplas}}, \bibinfo {author} {\bibfnamefont {A.}~\bibnamefont {Passos}}, \bibinfo {author} {\bibfnamefont {D.}~\bibnamefont {Cournapeau}}, \bibinfo {author} {\bibfnamefont {M.}~\bibnamefont {Brucher}}, \bibinfo {author} {\bibfnamefont {M.}~\bibnamefont {Perrot}},\ and\ \bibinfo {author} {\bibfnamefont
  {E.}~\bibnamefont {Duchesnay}},\ }\href@noop {} {\bibfield  {journal} {\bibinfo  {journal} {J. Mach. Learn. Res.}\ }\textbf {\bibinfo {volume} {12}},\ \bibinfo {pages} {2825} (\bibinfo {year} {2011})}\BibitemShut {NoStop}%
\bibitem [{\citenamefont {Gates}\ and\ \citenamefont {Ahn}(2019)}]{clusim}%
  \BibitemOpen
  \bibfield  {author} {\bibinfo {author} {\bibfnamefont {A.~J.}\ \bibnamefont {Gates}}\ and\ \bibinfo {author} {\bibfnamefont {Y.-Y.}\ \bibnamefont {Ahn}},\ }\href {https://doi.org/10.21105/joss.01264} {\bibfield  {journal} {\bibinfo  {journal} {J. Open Source Softw.}\ }\textbf {\bibinfo {volume} {4}},\ \bibinfo {pages} {1264} (\bibinfo {year} {2019})}\BibitemShut {NoStop}%
\bibitem [{\citenamefont {Peel}\ \emph {et~al.}(2017)\citenamefont {Peel}, \citenamefont {Larremore},\ and\ \citenamefont {Clauset}}]{GroundTruth}%
  \BibitemOpen
  \bibfield  {author} {\bibinfo {author} {\bibfnamefont {L.}~\bibnamefont {Peel}}, \bibinfo {author} {\bibfnamefont {D.~B.}\ \bibnamefont {Larremore}},\ and\ \bibinfo {author} {\bibfnamefont {A.}~\bibnamefont {Clauset}},\ }\bibfield  {journal} {\bibinfo  {journal} {Sci. Adv.}\ }\textbf {\bibinfo {volume} {3}},\ \href {https://doi.org/10.1126/sciadv.1602548} {10.1126/sciadv.1602548} (\bibinfo {year} {2017})\BibitemShut {NoStop}%
\bibitem [{\citenamefont {Tandon}\ \emph {et~al.}(2019)\citenamefont {Tandon}, \citenamefont {Albeshri}, \citenamefont {Thayananthan}, \citenamefont {Alhalabi},\ and\ \citenamefont {Fortunato}}]{fast_consensus}%
  \BibitemOpen
  \bibfield  {author} {\bibinfo {author} {\bibfnamefont {A.}~\bibnamefont {Tandon}}, \bibinfo {author} {\bibfnamefont {A.}~\bibnamefont {Albeshri}}, \bibinfo {author} {\bibfnamefont {V.}~\bibnamefont {Thayananthan}}, \bibinfo {author} {\bibfnamefont {W.}~\bibnamefont {Alhalabi}},\ and\ \bibinfo {author} {\bibfnamefont {S.}~\bibnamefont {Fortunato}},\ }\href {https://doi.org/10.1103/PhysRevE.99.042301} {\bibfield  {journal} {\bibinfo  {journal} {Phys. Rev. E}\ }\textbf {\bibinfo {volume} {99}},\ \bibinfo {pages} {042301} (\bibinfo {year} {2019})}\BibitemShut {NoStop}%
\bibitem [{\citenamefont {Rossi}\ and\ \citenamefont {Ahmed}(2022)}]{radoslaw}%
  \BibitemOpen
  \bibfield  {author} {\bibinfo {author} {\bibfnamefont {R.}~\bibnamefont {Rossi}}\ and\ \bibinfo {author} {\bibfnamefont {N.}~\bibnamefont {Ahmed}},\ }\href@noop {} {\bibinfo {title} {Ia-radoslaw-email}},\ \bibinfo {howpublished} {https://networkrepository.com/ia-radoslaw-email.php} (\bibinfo {year} {Last Accessed: June 14, 2022})\BibitemShut {NoStop}%
\bibitem [{\citenamefont {Kunegis}(2022)}]{enron}%
  \BibitemOpen
  \bibfield  {author} {\bibinfo {author} {\bibfnamefont {J.}~\bibnamefont {Kunegis}},\ }\href@noop {} {\bibinfo {title} {Enron}},\ \bibinfo {howpublished} {http://konect.cc/networks/enron} (\bibinfo {year} {Last Accessed: July 13, 2022})\BibitemShut {NoStop}%
\bibitem [{\citenamefont {Leskovec}(2022)}]{EUemail}%
  \BibitemOpen
  \bibfield  {author} {\bibinfo {author} {\bibfnamefont {J.}~\bibnamefont {Leskovec}},\ }\href@noop {} {\bibinfo {title} {email-eu-core temporal network}},\ \bibinfo {howpublished} {http://snap.stanford.edu/data/email-Eu-core-temporal.html} (\bibinfo {year} {Last Accessed: June 15, 2022})\BibitemShut {NoStop}%
\end{thebibliography}%

\end{document}